\documentclass[%
reprint,
superscriptaddress,
amsmath,amssymb, 
aps,
article,pre,
]{revtex4-2}

\usepackage{ulem,cancel}
\usepackage{graphicx}% Include figure files
\usepackage{dcolumn}% Align table columns on decimal point
\usepackage{bm}% bold math

\usepackage[T1]{fontenc}
\usepackage{mathptmx}
\usepackage{etoolbox}

\usepackage{enumerate}

\usepackage{cancel}
\usepackage{color,ulem}
\usepackage[colorlinks=true,citecolor=black,linkcolor=black,urlcolor=blue]{hyperref}
\usepackage{physics}

\newcolumntype{P}[1]{>{\centering\arraybackslash}p{#1}}
\newcommand{\mbf}{m_{\text{bf}}}

\newcommand{\st}{\text{s}}
\newcommand{\hcs}{\text{HCS}}
\newcommand{\LLNES}{\text{L}}
\newcommand{\Frz}{\text{Frz}}
\newcommand{\BL}{\text{(0)}}
\newcommand{\crit}{\text{crit}}

\begin{document}

\title{{Kinetic} glass transition in  granular gases and nonlinear molecular fluids}

\author{A. Patrón}%
 \email{apatron@us.es}
 \affiliation{Física Teórica, Universidad de Sevilla, Apartado de
  Correos 1065, E-41080 Sevilla, Spain}
\author{B. Sánchez-Rey}%
 \email{bernardo@us.es}
 \affiliation{Departamento de Física Aplicada I, Escuela Politécnica Superior, Universidad de Sevilla,  E-41011 Sevilla, Spain}
 \author{A. Prados}%
\email{prados@us.es}
\affiliation{Física Teórica, Universidad de Sevilla, Apartado de
  Correos 1065, E-41080 Sevilla, Spain}

\begin{abstract}
In this paper we investigate{, both analytically and numerically,} the emergence of a {kinetic} glass transition in two different {model} systems: a uniformly heated granular gas and a molecular fluid with nonlinear drag. Despite the profound differences between {these two physical} systems, their behavior in thermal cycles share strong similarities{, which stem from the relaxation time diverging algebraically at low temperatures for both systems.} When the driving intensity----for the granular gas---or the bath temperature---for the molecular fluid---is decreased to sufficiently low values, the kinetic temperature of both systems becomes ``frozen" at a value that depends on the cooling rate through a power law with the same exponent. Interestingly, this frozen glassy state is universal in the following sense: for a suitable rescaling of the relevant variables, its velocity distribution function becomes independent of the cooling rate. Upon reheating, i.e., when either the driving intensity or the bath temperature is increased from this frozen state, hysteresis cycles arise and the apparent heat capacity displays a maximum. {The numerical results obtained from the simulations are well described by a perturbative approach.}
\end{abstract}

%\pacs{Valid PACS appear here}% PACS, the Physics and Astronomy
                             % Classification Scheme.
%\keywords{Suggested keywords}%Use showkeys class option if keyword
                              %display desired
\maketitle
%tableofcontents
%\newpage

\section{Introduction}
As is well known, {crystallization of most liquids can be prevented by cooling them sufficiently fast.} In that case, the liquid enters into a metastable supercooled regime in which a dramatic slowing down of the dynamics takes place. On the one hand, above the melting point $T_m$, density fluctuations of the liquid relax on a time scale of the order of picoseconds. On the other hand, in the supercooled regime, the relaxation times increase so fast that they become 14 orders of magnitude larger when the temperature is around $\frac{2}{3}T_m$~\cite{debenedetti_supercooled_2001}. At this point, the liquid does not flow anymore and the glass transition occurs: configurational rearrangements cease, the liquid structure becomes ``frozen"  and the system  gets trapped in a nonequilibrium disordered yet solid state, called the glassy state~\cite{angell_formation_1995,dyre_colloquium_2006,berthier_theoretical_2011,wolynes_structural_2012,hunter_physics_2012,biroli_perspective_2013,stillinger_glass_2013,lubchenko_theory_2015,bomont_reflections_2017,weeks_introduction_2017,dauchot_glass_2022,novikov_temperature_2022,barrat_computer_2023,berthier_modern_2023}.

In spite of the great effort devoted to the investigation of glassy systems in the last decades, the glass transition continues to be an open problem. There is not yet a conclusive answer to the fundamental question of whether the glass transition is a purely dynamical, {kinetic}, phenomenon or is the consequence of an underlying {ideal} phase transition---as predicted in certain theoretical frameworks~\cite{berthier_theoretical_2011,wolynes_structural_2012,stillinger_glass_2013,bomont_reflections_2017,dauchot_glass_2022,novikov_temperature_2022}. Numerous studies have addressed the rich phenomenology that {accompanies} the glass transition from different and complementary viewpoints. For instance, spin models put the accent on the characterization of potential energy landscapes with a large number of energy minima connected by complex dynamics pathways~\cite{nishikawa_relaxation_2022,ros_complex_2019,ediger_glass_2021}, kinetically constrained models emphasize the fact that relaxation events are cooperative because of the presence of geometric frustration~\cite{ritort_glassy_2003,tarjus_frustration-based_2005}, and so on. But the development of a successful theory to explain all the phenomenological observations in a unified and satisfactory manner is still a challenge.

There are some key behaviors that are displayed by glass formers when submitted to cooling protocols followed by reheating. In the following, we exemplify the observed behavior with the average energy $\expval{E}$, but other physical quantity might be the relevant one---{depending on the context, for example} the average volume for polymeric glasses~\cite{kovacs_transition_1963,kovacs_isobaric_1979}. When the system is cooled down to a low temperature, e.g., by lowering the bath temperature $T$ at a constant rate $r_c$, the average energy $\expval{E}$ departs from equilibrium and gets frozen when the system relaxation time $\tau$ exceeds the characteristic cooling time $r_c^{-1}$. This purely kinetic phenomenon is termed the kinetic glass transition~\footnote{{We prefer to avoid the terminology ``laboratory glass transition'', which is also often employed, since our study is not experimental but analytical/computational.}}. The temperature of the {kinetic} glass transition---actually a range of temperatures---at which the system departs from equilibrium and gets frozen  decreases with the cooling rate and, consequently, the properties of a glass depend on the process by which it is formed. When the system is reheated from the frozen state {at a rate $r_h$, in general different from $r_c$,} $\expval{E}$ overshoots the equilibrium curve before returning thereto. This entails that the apparent~\footnote{We employ the term \textit{apparent} because the system departs from equilibrium and therefore $d\expval{E}/dT$ is here a dynamical quantity, depending on the rate of variation of the temperature, and it is not equal to the thermodynamic heat capacity.} heat capacity $d\expval{E}/dT$ displays a nontrivial behavior with a marked peak at a certain temperature $T_g$, which can be employed to characterize the {kinetic} glass transition~\cite{angell_formation_1995,dyre_colloquium_2006,gao_calorimetric_2013,tropin_modern_2016,richet_thermodynamics_2021}.

{This work aims at analyzing how the kinetic glass transition emerges}  in two specific {model} systems: a uniformly heated granular gas~\cite{van_noije_velocity_1998,van_noije_randomly_1999,montanero_computer_2000,garcia_de_soria_energy_2009,garcia_de_soria_universal_2012,prados_kovacs-like_2014} and a molecular fluid with nonlinear drag~\cite{klimontovich_nonlinear_1994,ferrari_particles_2007,ferrari_particles_2014,hohmann_individual_2017,santos_mpemba_2020,patron_strong_2021,megias_thermal_2022}. {This analysis is carried out by a combination of numerical simulations and singular perturbation theory tools. Granular gases and nonlinear molecular fluids} are largely different from a fundamental point of view. In the molecular fluid with nonlinear drag, collisions between particles are elastic and energy is thus conserved. Therefore, the nonlinear molecular fluid tends in the long-time limit to an equilibrium state, with a Gaussian---or Maxwellian---velocity distribution function (VDF). In the granular gas, collisions between particles are inelastic and thus energy is continuously lost. Therefore, an energy injection mechanism is necessary to drive the system to a steady state. The simplest one is the so-called stochastic thermostat, in which a stochastic forcing homogeneously acts on all the particles. In this uniformly heated granular gas, the system remains spatially homogeneous and tends in the long-time limit to a nonequilibrium steady state (NESS), in which the kinetic temperature is a certain function of the driving intensity. Moreover, the stationary VDF has a non-Gaussian shape, which is well described by the so-called first Sonine approximation. Therein, the non-Gaussianities are accounted for by the excess kurtosis, which is a smooth function of the inelasticity but independent of the driving intensity~\cite{van_noije_velocity_1998,montanero_computer_2000}.

Despite their apparent dissimilarities, uniformly heated granular systems and nonlinear molecular fluids share some features and characteristic behaviors. In both systems, { the system's Hamiltonian has only the kinetic contribution, since interactions proceed via hard-core collisions.} Therefore, the so-called kinetic temperature $T(t)$ univocally determines the average energy $\expval{E}(t)$, $T(t)\propto \expval{E}(t)$. Notwithstanding, the two systems display aging and associated memory effects~\cite{patron_nonequilibrium_2023}, both the Kovacs~\cite{prados_kovacs-like_2014,trizac_memory_2014,patron_strong_2021,patron_nonequilibrium_2023} and the Mpemba~\cite{lasanta_when_2017,patron_strong_2021,patron_nonequilibrium_2023} memory effects. The Kovacs memory effect is especially characteristic of the complex response of glassy systems~\cite{kovacs_transition_1963,kovacs_isobaric_1979,bertin_kovacs_2003,mossa_crossover_2004,arenzon_kovacs_2004,aquino_kovacs_2006,prados_kovacs_2010,bouchbinder_nonequilibrium_2010,lulli_kovacs_2021,godreche_glauber-ising_2022}. It is interesting to note that the Mpemba effect has also been observed in spin glasses, but only in the spin glass phase---where it arises due to the aging dynamics of the internal energy~\cite{baity-jesi_mpemba_2019}.

In addition, when quenched to a very low temperature, both granular gases and nonlinear molecular fluids tend to a time-dependent, nonequilibrium state, in which the kinetic temperature presents a very slowly nonexponential, algebraic, decay over a wide intermediate time window. These nonequilibrium attractors, the homogeneous cooling state (HCS) for the granular gas~\cite{brey_homogeneous_1996,poschel_granular_2001,garzo_granular_2019}
and the long-lived nonequilibrium state (LLNES) for the molecular fluid~\cite{patron_strong_2021,patron_nonequilibrium_2024}, are characterized by non-Gaussian VDFs. Afterwards, for very long times, both systems approach their respective stationary states,  NESS and equilibrium state, for the granular and molecular cases, respectively.

Since nonexponential relaxation and memory effects are hallmarks of glassy behavior~\cite{kovacs_transition_1963,kovacs_isobaric_1979,angell_relaxation_2000,bertin_kovacs_2003,mossa_crossover_2004,arenzon_kovacs_2004,aquino_kovacs_2006,prados_kovacs_2010,bouchbinder_nonequilibrium_2010,lulli_kovacs_2021,keim_memory_2019,baity-jesi_mpemba_2019,morgan_glassy_2020,song_activation_2020,godreche_glauber-ising_2022,patron_nonequilibrium_2023}, it is natural to  pose the question as whether granular gases and nonlinear molecular fluids undergo a {kinetic} glass transition when being subjected to a continuous cooling program. Of course, these systems are not realistic models of glass-forming liquids but one of the most interesting features of glassy behavior is its ubiquity and universality: the glass transition is found in systems with typical length and time scales very different from molecular ones---such as colloidal suspensions and granular materials~\cite{dauchot_glass_2022}.
More specifically, we would also like to elucidate the possible role played by the HCS---for the granular gas---and the LLNES---for the nonlinear molecular fluid---in the {kinetic} glass transition.

The organization of the paper is as follows. In Sec.~\ref{section2} we introduce the uniformly heated granular gas model and write the evolution equations for the kinetic temperature and the excess kurtosis---in the first Sonine approximation that we employ {throughout} our work.
{ Section~\ref{sec:why-kinetic-glass-transition} analyses the physical reasons behind the emergence of a  kinetic glass transition in the uniformly heated granular gas when the driving intensity is continuously decreased to zero. For the sake of concreteness, we consider a linear cooling program, in which the bath temperature changes linearly in time. The divergence of the characteristic relaxation time as the bath temperature is lowered gives rise to a slowing down of the dynamics that makes the granular temperature become frozen at a certain value $\theta^\Frz \ne 0$, and physical arguments for the dependence of $\theta^\Frz$ on $r_c$ are provided. In Sec.~\ref{sec:math-descr-glass},  we analytically } investigate the {kinetic} glass transition in the granular gas. Not only do we perform numerical simulations of the system under this cooling program but also develop a singular perturbation theory approach---specifically, of boundary layer type, {which accurately} accounts for the system evolution and even characterizes the final glassy state. The hysteresis cycle that emerges when the system is reheated from {the} final glassy, frozen, state  is the subject of Sec.~\ref{section4}. The molecular fluid with nonlinear drag model is introduced in Sec.~\ref{section5}, where---similarly to the framework developed in Sec.~\ref{section2} for the granular gas---the evolution equations of the model in the first Sonine approximation are put forward. In Sec.~\ref{section-glass-molecular}, we address the glass transition and hysteresis cycles in {the molecular fluid}, by combining  again numerical simulations and a boundary layer approach---this analysis is presented in a simplified way, because of its formal similarity with the granular gas. Finally, we present in Sec.~\ref{section6} the main conclusions and a brief discussion of our results. The appendices {present the study of} more general cooling programs and  give additional details on the {boundary layer} perturbation theory { employed at different points of our work.}

\section{Model: Uniformly driven granular gas}\label{section2}

First, we consider a granular gas of $d$-dimensional hard spheres of mass $m$ and diameter $\sigma$. These hard spheres undergo binary inelastic collisions, in which the tangential component of the relative velocity between two particles remains unaltered, while the normal component is reversed and shrunk by a factor $\alpha$. This parameter $\alpha$ is called the restitution coefficient, $0\le \alpha \le 1$; elastic collisions---in which the kinetic energy is conserved---are recovered for $\alpha=1$~\cite{poschel_granular_2001,garzo_granular_2019}. In the uniformly heated granular gas, the system reaches a steady state in the long term because the kinetic energy lost in collisions is balanced on average by energy inputs, modeled through independent white noise forces acting  over each particle~\cite{van_noije_velocity_1998}.

{For the sake of clarity, and also to keep our work as self-contained as possible, we briefly present the general mathematical framework and the basic equations for the uniformly heated granular gas below. This framework is given for (i) sufficiently dilute gases and (ii) spatially homogeneous and isotropic situations. {On the one hand, (i) implies that the state of the system is completely characterized at the one-particle level, such that its dynamical evolution is accounted by a Boltzmann-like kinetic equation for the one-particle distribution function $f(\bm{r},\bm{v},t)$, and, on the other hand, (ii) allows to further simplify the description, as the distribution function becomes independent of $\bm{r}$. In the following, we refer to $f(\bm{v},t)$ as the one-particle VDF.} A more detailed account { of the mathematical framework described below} can be found in the literature---e.g., see Refs.~\onlinecite{van_noije_velocity_1998,van_noije_randomly_1999,montanero_computer_2000,garcia_de_soria_universal_2012,prados_kovacs-like_2014,trizac_memory_2014}.}

{In kinetic theory~\cite{resibois_classical_1977,chapman_mathematical_1990}, the one-particle VDF is usually normalized as
\begin{equation}
    n = \int d\bm{v}f(\bm{v},t),
\end{equation}
with $n$ being the number density, i.e., the number of particles per unit volume. The average of any function of the velocity $h(\bm{v})$ is
\begin{equation}
    \expval{h(\bm{v})}=\frac{\int d\bm{v}\ h(\bm{v}) f(\bm{v},t)}{\int d\bm{v} f(\bm{v},t)}=\frac{1}{n}\int d\bm{v}\ h(\bm{v}) f(\bm{v},t).
\end{equation}
The time evolution of the VDF is governed by the Boltzmann equation with an additional Fokker-Planck term, known as the Boltzmann-Fokker-Planck equation:}
\begin{equation}\label{eq:BFPE}
    \partial_t f(\bm{v},t) - \frac{\xi^2}{2}\frac{\partial^2}{\partial \bm{v}^2}f(\bm{v},t) = J_{\alpha}[\bm{v}|f,f].
\end{equation}
{On the one hand, the Boltzmann collision operator} $J_{\alpha}[\bm{v}|f,f]$ accounts for the inelastic collisions between the particles. We do not provide its full expression, {since our approach starts from the evolution equations for the cumulants, written below~\footnote{Further details on this integral operator may be found, for instance, in Refs.~\onlinecite{poschel_granular_2001,garzo_granular_2019}.}. On the other hand, the Fokker-Planck term accounts for the effect of the stochastic thermostat, with the parameter $\xi$ measuring the intensity of the ``heating''.} 

The kinetic (or granular) temperature $T(t)$ is defined as usual {in kinetic theory}, proportional to the average kinetic energy:
\begin{equation}
\label{suppl-temperature-gran}
    T(t) = \frac{m}{dk_B}\left\langle  v^2\right\rangle ,
\end{equation}
where $k_B$ is the Boltzmann constant. In order to gain analytical insights on the evolution of the granular temperature, it is useful to introduce the scaled VDF $\phi (\bm{c},t)$
\begin{equation}\label{eq:scaled-variables}
    f(\bm{v},t) = \frac{n}{v_T^d(t)}\phi(\bm{c},t), \quad \bm{c}\equiv  \frac{\bm{v}}{v_T(t)},
\end{equation}
with $v_T(t) \equiv \sqrt{2k_BT(t)/m}$ being the thermal velocity. For isotropic states, such scaled VDF may be expanded in a complete set of orthogonal polynomials: 
\begin{equation}
\label{eq:sonine}
    \phi (\bm{c},t) = \frac{e^{-c^2}}{\pi^{d/2}}\left[1+\sum_{l=2}^{\infty} a_l(t)\, L_l^{\frac{d-2}{2}}(c^2)\right].
\end{equation}
{Herein,} $L_l^{(k)}$ are the Sonine polynomials~\cite{goldshtein_mechanics_1995,poschel_granular_2001,santos_second_2009,garzo_granular_2019}, and the $a_l(t)$ coefficients are known as the Sonine cumulants. The latter account for the deviations from the Maxwellian equilibrium distribution $\phi_{\text{eq}}(\bm{c})=\pi^{-d/2} e^{-c^2}$.  

Throughout this work, here for the granular gas---and later for the molecular fluid, we work under the first Sonine approximation. Therein, we only need to monitor the kinetic temperature $T$ and the first Sonine cumulant $a_2$, given by
\begin{equation}
    \label{suppl-excess-kurtosis}
    a_2 = \frac{d}{d+2}\frac{\left\langle v^4\right\rangle}{\left\langle v^2\right\rangle^2}-1,
\end{equation}
which is also known as the excess kurtosis. For our analysis below, it is useful to introduce a characteristic length $\lambda$ and a characteristic rate $\nu$ as
\begin{subequations}
  \begin{align}
    \lambda^{-1}=&\frac{2 n \sigma^{d-1}  \pi^{\frac{d-1}{2}}}{d \,\Gamma(d/2)}, \\ 
    \label{subeq:coll-rate}
    \nu(T)=&\left(1-\alpha^2\right)\lambda^{-1} \left(\frac{k_B T}{m}\right)^{1/2}.
\end{align}  
\end{subequations}
{ On the one hand, $\lambda$ gives the mean free path, i.e. the average distance travelled by one particle between collisions. On the other hand, $\nu(T)$ gives the cooling rate of the granular gas, i.e. the rate at which kinetic energy is dissipated in collisions.}

In the absence of stochastic thermostat, the granular gas reaches the spatially-uniform nonsteady state known as the HCS, for which the scaled VDF $\phi$ becomes stationary and the granular temperature decays algebraically in time, $T(t) \propto t^{-2}$, following Haff's law~\cite{haff_grain_1983,brey_homogeneous_1996,poschel_granular_2001,garzo_granular_2019}. Under the first Sonine approximation, the stationary value of the excess kurtosis at the HCS is given by 
\begin{equation} 
    \label{eq:a2hcs}
    a_2^{\hcs}=\frac{16(1-\alpha)(1-2\alpha^2)}{25+2\alpha^2 (\alpha-1) +24 d+\alpha (8d-57)}.
\end{equation}

When the stochastic thermostat is present, the granular gas reaches a NESS in the long time limit. The { kinetic} temperature $T_{\st}$ at the NESS is given in terms of the stochastic strength $\xi$ via the relation~\cite{van_noije_velocity_1998}
\begin{equation}
    \frac{k_B T_{\st}}{m} = \left[\frac{\lambda\, \xi^2}{\left(1-\alpha^2\right)\left(1+\frac{3}{16}a_2^{\st}\right)}\right]^{2/3},
    \label{eq:Ts-xi}
\end{equation}
where $a_2^{\st}$ is the NESS value of the excess kurtosis,
\begin{equation} \label{eq:a2s}
  a_2^\st=\frac{16(1-\alpha)(1-2\alpha^2)}
{73+56d-24d\alpha-105\alpha+30(1-\alpha)\alpha^2}.
\end{equation}
Such value has the same sign as $a_2^{\hcs}$, thus attaining a null value at $\alpha = 1/\sqrt{2}$.

From the Boltzmann-Fokker-Planck equation, the evolution equations for the temperature and the excess kurtosis are derived~\cite{van_noije_velocity_1998,montanero_computer_2000,prados_kovacs-like_2014,trizac_memory_2014},
\begin{subequations}\label{granularSonine}
\begin{align}
    \frac{d\theta}{dt^*}&=  \theta_{\st}^{3/2} \left(1+\frac{3}{16} a_2^{\st}\right) - \theta^{3/2} \left(1+\frac{3}{16} a_2\right),
    \label{granularSonine-theta}\\
    \frac{da_2}{dt^*}&= 2\theta^{1/2} \left[  \left(1-\left(\frac{\theta_{\st}}{\theta}\right)^{3/2} \right) a_2 +B \, (a_2^{\st} -a_2) \right],
    \label{granularSonine-a2}
\end{align}
\end{subequations}
where we have {introduced dimensionless variables,}
\begin{equation}
	\label{adim_var}
	\theta \equiv
	\frac{T}{T_i}  , \quad \theta_{\st} \equiv \frac{T_{\st}}{T_i}, \quad t^*\equiv \nu(T_i)t,
\end{equation}
{ being  $T_i \equiv T (t = 0)$ the initial temperature,} and  the parameter
\begin{align}\label{eq:B-param}
B &\equiv \frac{73+8d(7-3\alpha)+15\alpha[2\alpha(1-\alpha)-7]}
{16(1-\alpha)(3+2d+2\alpha^2)}.
\end{align}
With our choice of units, the initial value of the dimensionless temperature is always $\theta_i=1$. Also, note that $B$ may be written in terms of $a_2^{\hcs}$ and $a_2^{\st}$, specifically $B = a_2^{\hcs}/(a_2^{\hcs}-a_2^{\st})$ as predicted by Eq.~\eqref{granularSonine-a2} for $\theta_{\st}=0$~\cite{prados_kovacs-like_2014,trizac_memory_2014}.

In order to simplify our notation, we drop the asterisk in the dimensionless time in the remainder of the paper. { Our definition of the dimensionless time is equivalent to taking the time unit roughly equal to the characteristic relaxation time of the granular gas at the initial temperature.}

{
\section{Why a kinetic glass transition in the granular gas?}\label{sec:why-kinetic-glass-transition}

The possible emergence of a glass transition in a given system is deeply connected with a slowing down of its dynamics, typically as the bath temperature is lowered. In the granular gas, the role of the heat bath is played by the intensity of the stochastic thermostat $\xi$, which controls the stationary value of the energy of the gas---or, equivalently, the stationary value of the kinetic temperature $\theta_{\st}$, as given by Eq.~\eqref{eq:Ts-xi}. A time-dependent driving intensity $\xi(t)$, which continuously decreases from its initial value $\xi_i$ to zero, is thus considered. 

The system is initially prepared in the NESS corresponding to $\xi_i$, thus the initial value of the dimensionless temperature is $\theta_i=\theta(t=0) = \theta_{\st}(t=0) = 1$. Therefrom, we apply a linear cooling program with rate $r_c$,
\begin{equation}
\label{cooling-programme}
 \frac{d\theta_{\st}}{dt} = -r_c, \quad \theta_{\st}(t)=1-r_c\, t.
\end{equation}
The choice of a linear cooling program is done for the sake of concreteness, but a more general family of protocols is considered in Appendix~\ref{sec:appendix-2}. The characteristic timescale for the cooling process $t_0$ corresponds to the time at which $\theta_{\st}$ vanishes, $\theta_{\st}(t=t_0)=0$---for the linear cooling program above, $t_0=r_c^{-1}$. The cooling process is assumed to be slow, i.e. the characteristic cooling time $t_0$ is much longer than the characteristic relaxation time of the system at the initial state. Since the latter is of the order of unity in our dimensionless variables, slow cooling entails that $t_0\gg 1$ or $r_c\ll 1$.

Now we put forward a physical argument that supports the emergence of a kinetic glass transition in the granular gas. For time-independent strength of the stochastic thermostat $\xi$, the granular gas relaxes to the steady state characterized by the ``bath" temperature $\theta_{\st}$ and the stationary excess kurtosis $a_2^{\st}$. The characteristic timescale for this relaxation is is proportional to $\nu^{-1}(T_{\st})$ from Eq.~\eqref{subeq:coll-rate}. Specifically, it is given by 
\begin{equation}\label{eq:tau-def}
    \tau(\theta_{\st})= \frac{1}{c}\theta_{\st}^{-1/2},
\end{equation}
where $c$ is a constant of the order of unity~\cite{sanchez-rey_linear_2021}---see Appendix~\ref{app:linear-relaxation-time}---which is approximately equal to $3/2$. In the low bath temperature limit, $\tau$ algebraically diverges as $\theta_{\st}^{-1/2}$  and, despite our slow cooling, the characteristic timescale for relaxation eventually becomes much longer than the cooling time. Therefore, we expect the system to depart from the stationary curve and get frozen---i.e. a kinetic glass transition shows up.

In order to approximately quantify the above qualitative argument, we may introduce the effective timescale
\begin{equation}\label{eq:s-scale-def}
    s=\int_{t}^{t_0}dt' \tau^{-1}(\theta_{\st}(t'))=\frac{1}{r_c}\int_{0}^{\theta_{\st}} d\theta_{\st}' \, \tau^{-1}(\theta_{\st}') ,
\end{equation}
which measures the number of effective relaxation times remaining from the current time $t$ to the final time of the cooling process $t_0$. As long as $s\gg 1$, we expect the system to be able to follow the instantaneous NESS curve $\theta=\theta_{\st}$~\footnote{Note that our defining of the timescale $s$ also allows for a more precise definition of slow cooling: a cooling program is said to be slow if $s(t=0)\gg 1$, i.e. the system has time to reach the instantaneous NESS curve before becoming frozen, independently of the initial preparation.}. When $s$ becomes of the order of unity, the system does not have enough time to relax towards the instantaneous NESS curve and thus it freezes. Following the usual terminology of glassy systems, see e.g.~\cite{scherer_relaxation_1986}, we may introduce a fictive kinetic temperature, as the bath temperature at which the NESS kinetic temperature equals the frozen value. 

The above physical picture implies that we can estimate the fictive temperature $\theta_f$ by imposing
\begin{equation}
    s(\theta_{\st}=\theta_f)=1,
\end{equation}
i.e.
\begin{equation}
    \theta^{\Frz}\equiv \lim_{\theta_{\st}\to 0} \theta \simeq \theta_f.
\end{equation}
Bringing to bear Eqs.~\eqref{cooling-programme} and \eqref{eq:tau-def},
\begin{equation}\label{eq:s-scale-v2}
    s=\frac{c}{r_c}\int_{0}^{\theta_{\st}}d\theta_{\st}' \, \sqrt{\theta_{\st}'}=\frac{2c}{3} \frac{\theta_{\st}^{3/2}}{r_c}.
\end{equation}
Then, the fictive temperature and the kinetic temperature at the frozen state are estimated as
\begin{equation}\label{eq:thetaf-power-law}
    \theta_f =\left(\frac{3r_c}{2c}\right)^{2/3} , \quad \theta_{\Frz}\propto r_c^{2/3}.
\end{equation}

Wrapping things up, the slowing down of the dynamics of the granular gas, due to the algebraic divergence of the relaxation time in Eq.~\eqref{eq:tau-def} entails that the granular gas is expected to depart from the instantaneous NESS curve $\theta=\theta_{\st}$ as the the intensity of the stochastic thermostat is continuously decreased to zero. In other words, a kinetic glass transition is expected to appear in this system when cooled down to low bath temperatures, and our timescale argument suggests that the system would get frozen for bath temperatures $\theta_{\st}\lesssim \theta_f$, where $\theta_f$ follows the power law~\eqref{eq:thetaf-power-law} with the cooling rate. Moreover, the kinetic temperature at the frozen state is expected to be approximately equal to $\theta_f$, thus following the same power law with the cooling rate. The correctness of this physical image is supported by the detailed mathematical theory that is developed in the next sections.

}

{
\section{Detailed analysis of the glass transition}\label{sec:math-descr-glass}
}

{ The physical discussion in the previous section} suggests that tools from singular perturbation theory are useful to tackle the problem analytically. Our analysis below shows that indeed different regions emerge, which we will label with the terminology of Ref.~\onlinecite{bender_advanced_1999} for boundary layer problems. First, one has the \textit{outer layer}, inside which the kinetic temperature $\theta$ does not deviate much from the bath temperature $\theta_{\st}$, and a regular perturbation expansion is adequate. Second, one has the \textit{inner layer}, for which the regular perturbation expansion breaks down and it is necessary to rescale the variables to obtain an approximate solution. It is in the inner layer that the kinetic temperature $\theta$ rapidly separates from $\theta_{\st}$ and gets frozen at $\theta^{\Frz}$. Finally, one has the \textit{matching region}, over which the solution continuously changes from the inner to the outer solution.

The generic framework described above is applied to the evolution equations for the granular gas in the following sections \ref{subsec:pert-mol} and \ref{sec:bl-universality}. { In order to improve their readability, some of the details of the perturbative approach are omitted or relegated to the appendices.} 
\\

\subsection{\label{subsec:pert-mol}Regular perturbative expansion}

{ We are interested in studying the behavior of the kinetic temperature $\theta$ in terms of the bath temperature $\theta_{\st}$. Therefore, we rewrite Eqs.~\eqref{granularSonine} in terms of derivatives with respect to $\theta_{\st}$  }
\begin{subequations}\label{granularSonine-v2}
\begin{align}
    -r_c \frac{d\theta}{d\theta_{\st}}&=  \theta_{\st}^{3/2} \left(1+\frac{3}{16} a_2^{\st}\right) - \theta^{3/2} \left(1+\frac{3}{16} a_2\right),
    \label{granularSonine-theta-v2}\\
    -r_c \frac{da_2}{d\theta_{\st}}&= 2\theta^{1/2} \left[  \left(1-\left(\frac{\theta_{\st}}{\theta}\right)^{3/2} \right) a_2 +B \, (a_2^{\st} -a_2) \right].
    \label{granularSonine-a2-vw}
\end{align}
\end{subequations}
{ which must be solved with the conditions}
\begin{equation}
    \theta(\theta_{\st}=1)=1, \quad a_2(\theta_{\st}=1)=a_2^{\st}.
    \label{eq:granular-Sonine-bc}
\end{equation}

In the slow cooling limit $r_c\ll 1$, 
{
a standard, regular, perturbative approach in powers of $r_c$---for details, see Appendix~\ref{app:reg-granular-gas}---gives the ``outer" solution
\begin{subequations}
\label{outer-gran-cooling}
	\begin{align}
		\theta &= \theta_{\st}+ r_c \frac{2}{3\theta_{\st}^{1/2}} \left[1+\frac{3}{16} \; a_2^{\st} \left( 1+\frac{1}{B} \right) \right]^{-1}+O(r_c^2), \label{outer-gran-cooling-temp}
			\\
		a_{2} &= a_2^{\st}-r_c \frac{\,a_2^{\st}}{\,B\,\theta_{\st}^{3/2}}  \left[1+\frac{3}{16} \; a_2^{\st} \left( 1+\frac{1}{B} \right) \right]^{-1}+O(r_c^2). 
	\end{align}
\end{subequations}
This expansion breaks down for low bath temperatures $\theta_{\st}\ll 1$, for which the terms proportional to $r_c$ (i) first become of the order of the leading, independent of $r_c$, contributions and (ii) later diverge in the limit as $\theta_{\st}\to 0$. In particular, (i) implies that Eq.~\eqref{outer-gran-cooling} ceases to be valid when $\theta_{\st} = O(r_c^{2/3})$. In other words, Eq.~\eqref{outer-gran-cooling} is thus limited to high enough bath temperatures, $\theta_{\st}\gg r_c^{2/3}$. 

Note that $r_c^{2/3}$ is precisely the dependence on $r_c$ of the fictive temperature $\theta_f$ derived in Sec.~\ref{sec:why-kinetic-glass-transition} by qualitative arguments.
}
% To the lowest order, we thus have the \textit{outer solution}
% \begin{equation}
%     \theta_{\text{O}} \equiv \theta^{(0)} = \theta_{\st}, \quad a_{2,\text{O}} \equiv a_2^{(0)} = a_2^{\st}, \qquad \theta_{\st}\gg r_c^{2/3}.
%     \label{eq:outer-sol}
% \end{equation}
% which has striking implications. As argued above, this regular expansion ceases to be valid for low enough bath temperatures, $\theta_{\st}=O(r_c^{2/3}$) or smaller. From a mathematical standpoint, this scaling gives the width of the boundary layer---see Sec.~\ref{sec:bl-universality} for further details. 
From a physical standpoint, this marks the onset of the kinetic glass transition: we expect the system to become ``frozen'' as soon as $\theta_{\st}=O(r_c^{2/3})$. In this region of temperatures, the terms in Eq.~\eqref{outer-gran-cooling} for (i) $\theta$ are proportional to $r_c^{2/3}$ and (ii) $a_2$ are independent of $r_c$. Therefore, we expect that
\begin{subequations}\label{eq:scaling-outer-v1}
 \begin{align}
     \theta^{\Frz}&\equiv \lim_{\theta_{\st}\to 0} \theta \propto r_c^{2/3} ;
    \label{freezescaling} \\
    \label{scaling-2}
    a_2^{\Frz} &\equiv \lim_{\theta_{\st}\to 0} a_2 = O(1),  
 \end{align} 
\end{subequations}
The latter suggests that all the cumulants of the Sonine expansion become independent of the cooling rate, i.e., the frozen state of the system is unique~\footnote{The scalings proposed in Eq.~\eqref{eq:scaling-outer-v1}, which arise from the above { qualitative arguments, are  quantitatively justified by the boundary layer approach carried out in Sec.~\ref{sec:bl-universality}.}}.

Let us now compare our analytical predictions  with simulation results obtained from Direct Simulation Monte Carlo (DSMC) integration~\cite{bird_g_a_molecular_1994} of the {Boltzmann} equation~\eqref{eq:BFPE} that governs the dynamics of the granular gas. Unless otherwise specified, for all of the simulations of the granular gas performed,  we have employed the system parameters $d=3$, a number of particles {of} $N=10^5$, {and two different values of the restitution coefficient: $\alpha=0.9$ and $\alpha=0.3$, in order to test the robustness of our theoretical approach.} 

\begin{figure*}
  \centering 
 {\centering \includegraphics[width=3.05in]{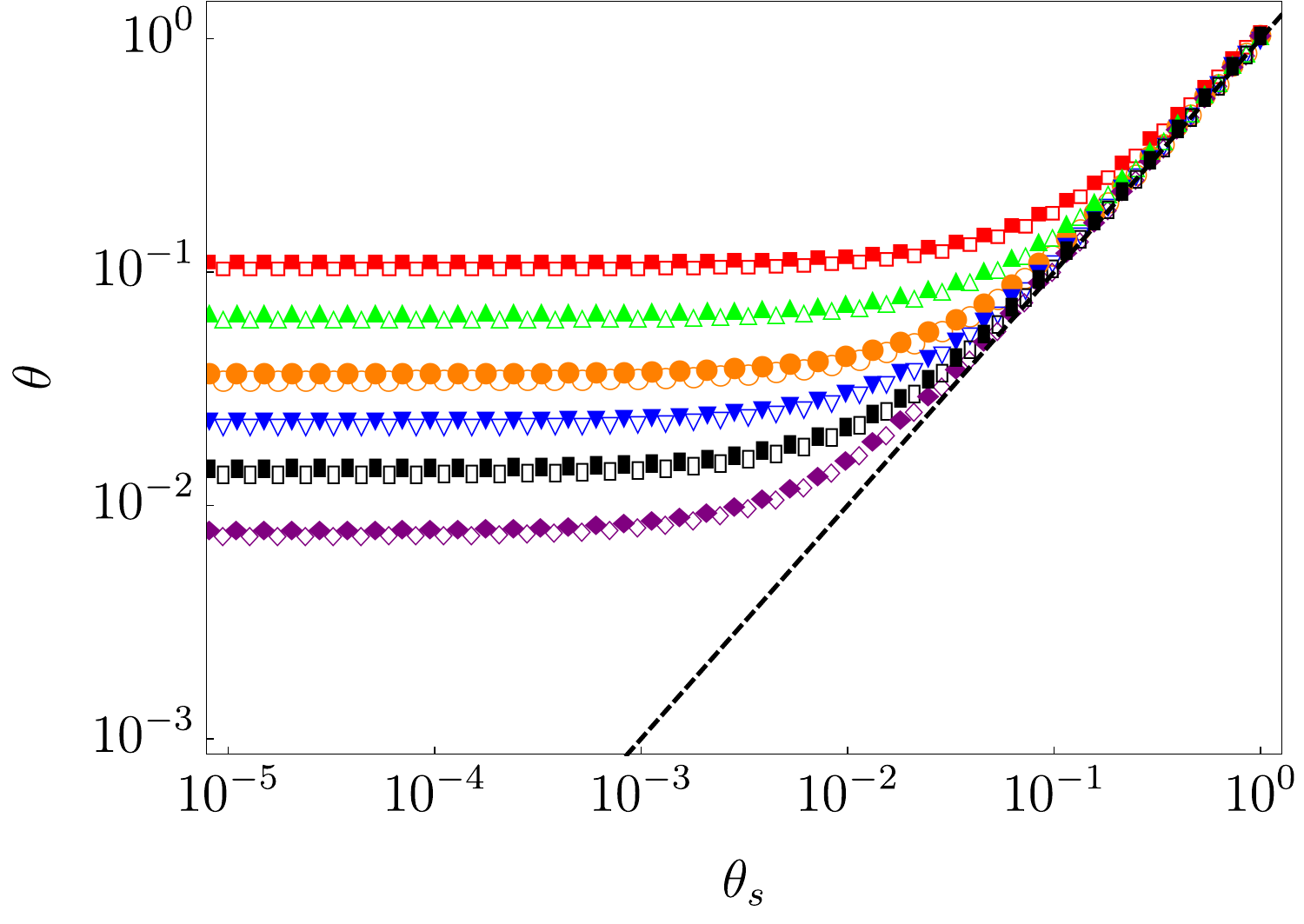}\hspace*{2em}
 \includegraphics[width=3.05in]{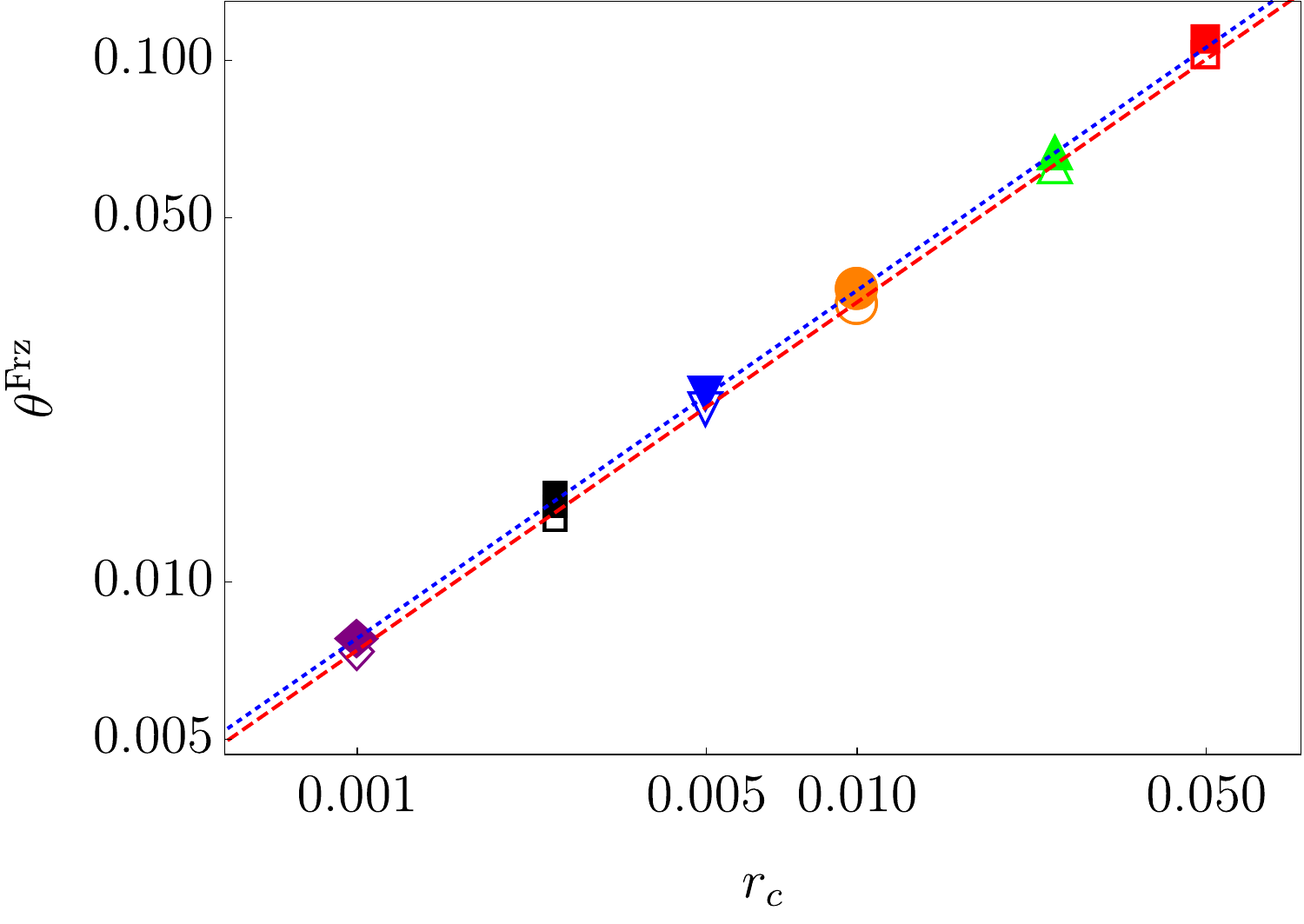}}
  \caption{(Left) Dynamical evolution of the granular temperature $\theta$ as a function of the bath temperature $\theta_{\st}$. {Symbols} correspond to DSMC  data for the linear cooling protocol \eqref{cooling-programme} with different {cooling} rates $r_c$, namely: $r_c$= 0.05 (red squares), $0.025$ (green up triangles), $0.01$ (orange circles), $0.005$ (blue down triangles), $0.0025$ (black rectangles), and $0.001$ (purple diamonds) {for two values of the restitution coefficient: $\alpha = 0.9$ (empty symbols)  and  $\alpha = 0.3$ (filled symbols)}. The dashed line corresponds to the instantaneous NESS curve $\theta = \theta_{\st}$. (Right)  Limit values of the kinetic temperature at the frozen state {$\theta^{\Frz}$} as a function of {$r_c$.} The plotted points have been extracted from the DSMC data on the left panel. { The red, dashed and blue, dotted lines correspond to the best fits to the function ${\theta^{\Frz}=}a\, r_c^{b}$, with $a = 0.741$ and $b = 0.666$ for $\alpha = 0.9$, and $a = 0.781$ and $b = 0.666$ for $\alpha = 0.3$, both} in excellent agreement with the theoretical prediction \eqref{freezescaling}. We have considered a granular gas in the three-dimensional case $d = 3$. The same parameter values are employed in the remainder of the numerical simulations for the granular gas.
  }
  \label{frozen-temp-gran}
\end{figure*}

{Figure~\ref{frozen-temp-gran} presents numerical results for the linear cooling program~\eqref{cooling-programme}. On the left panel, we plot the behavior of the kinetic temperature $\theta$ as a function of the bath temperature $\theta_{\st}$, for different cooling rates. The emergence of a kinetic glass transition is clearly observed. On the right panel, the final granular temperatures $\theta^{\Frz}$ at the frozen state are plotted as a function of the cooling rate $r_c$.} They are very well fitted  by the power law  $\theta^{\Frz}=a\, r_c^{b}$ with $b=0.666$, thus {numerically} confirming the scaling {predicted} by Eq.~\eqref{freezescaling}. 

\begin{figure*}
	\centering
	\includegraphics[width=3in]{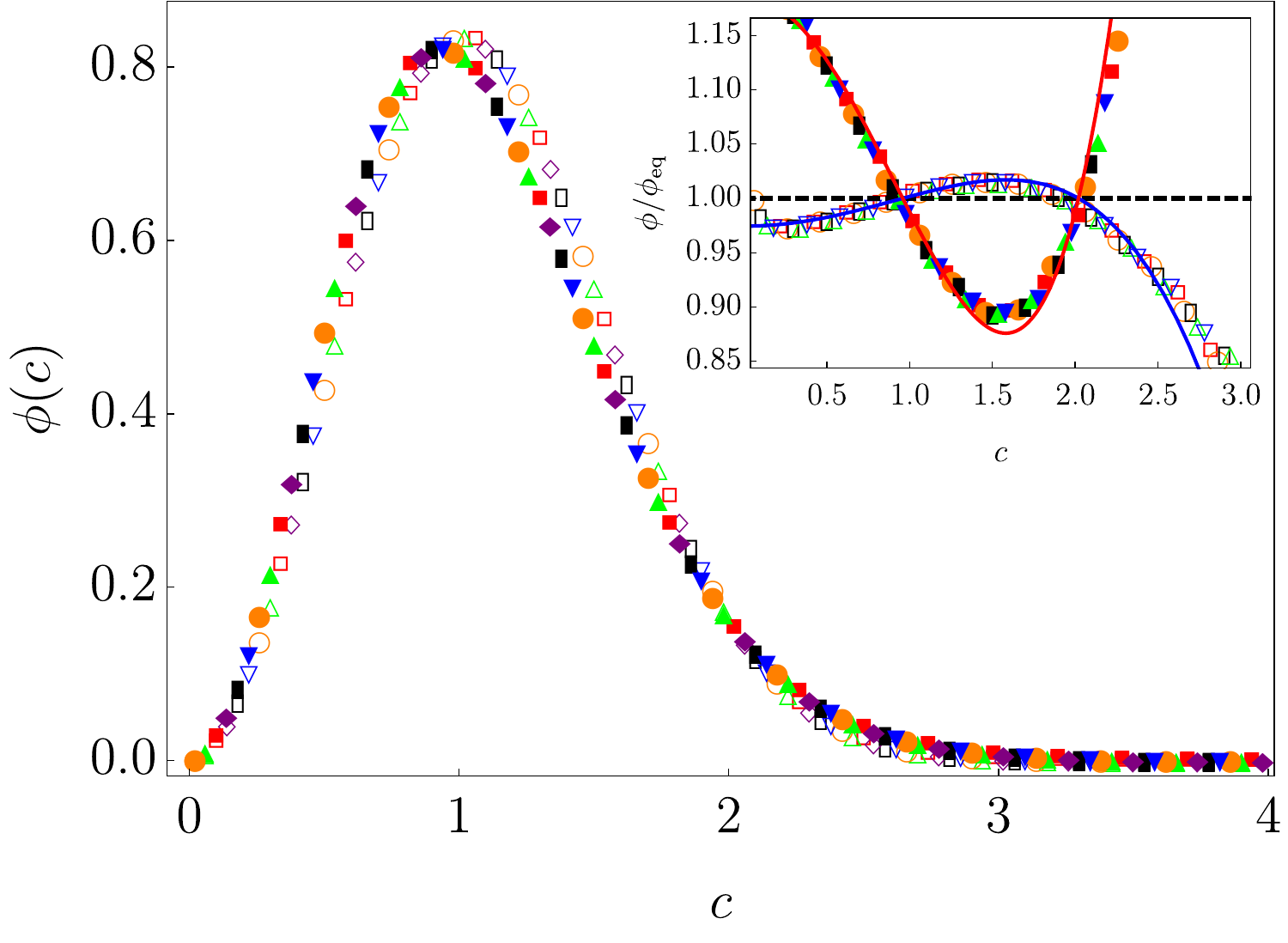} 
    \includegraphics[width=3in]{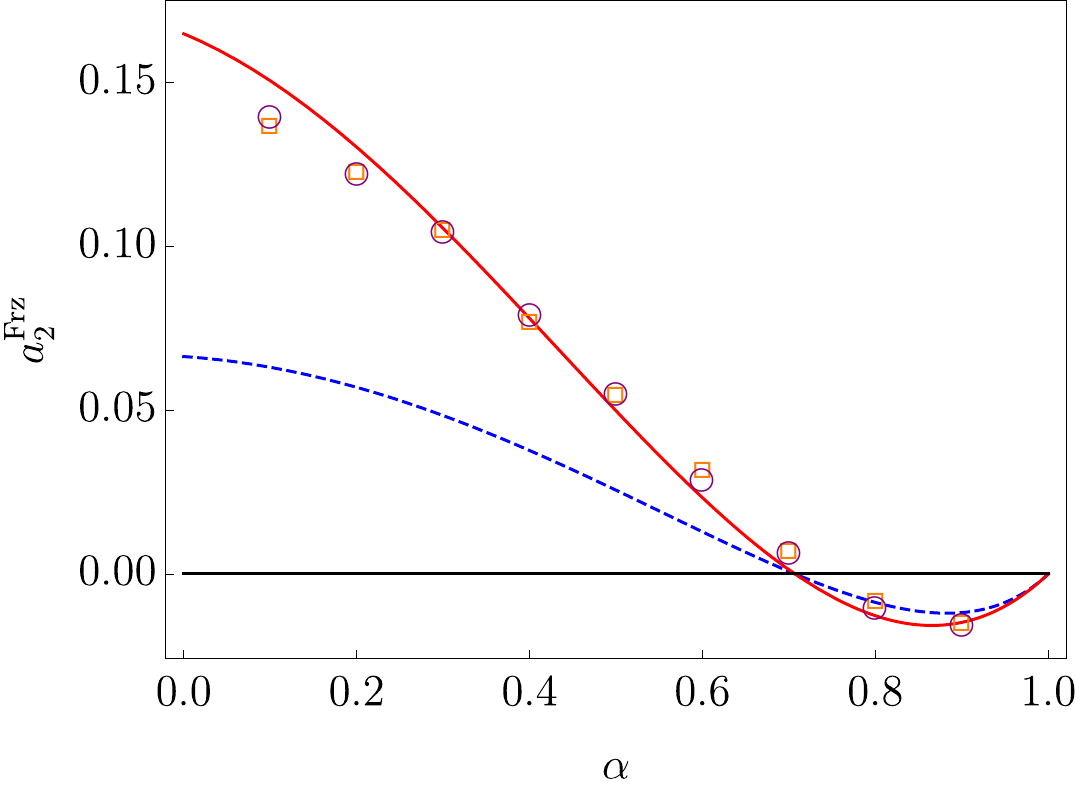} 
 \caption{Universality of the frozen state. (Left) VDF at the frozen state for different values of the cooling rate $r_c$. { The color code and symbols are the same as in Fig.~\ref{frozen-temp-gran}. For each value of $\alpha$, its corresponding} VDFs  are superimposed over a unique, universal, curve independent of $r_c$, in agreement with our theoretical prediction. In the inset, we show the VDF at the frozen state divided by the equilibrium Maxwellian, with the solid lines corresponding to the polynomials in Eq.~\eqref{eq:sonine} within the first Sonine approximation {for  $\alpha=0.9$ (blue) and  $\alpha=0.3$ (red), respectively.} {(Right) Excess kurtosis at the frozen state $a_2^{\Frz}$ as a function of the restitution coefficient $\alpha$. Here, for the sake of clarity, we show DSMC data corresponding to only two values of the cooling rate, $r_c=0.01$ (squares) and $r_c=0.001$ (circles). The numerical values {of $a_2^{\Frz}$} are compared with both the NESS value $a_2^{\st}$ (blue dashed line) and the HCS value $a_2^{\text{HCS}}$ (red solid line), being very close to the latter.} }
	\label{VDF-graph}
\end{figure*}

In Fig.~\ref{VDF-graph}, we {numerically} prove that the frozen state is indeed unique. On the left panel, the dimensionless VDFs at the frozen state corresponding to different cooling rates overlap on a universal curve. Note that, although our {theoretical argument for the universality of the frozen state above} has been carried out within the first Sonine approximation, the numerical results show that this remarkable property holds for the exact (numerical) VDF. To neatly visualize the non-Gaussian character of the frozen state, we present (i) the ratio of the VDFs over the equilibrium Maxwellian in the inset of the left panel and (ii) the excess kurtosis at the frozen state $a_2^{\Frz}$ as a function of the restitution coefficient $\alpha$ on the right panel~\footnote{In order to improve the statistics, we have employed a larger system with $N=10^6$ particles for both the inset and the right panel. In the former, the shown DSMC data has been further averaged over 10 trajectories.}. {Point (i) allows us to illustrate in a neater way the differences between the frozen states corresponding to $\alpha = 0.9$ and $\alpha = 0.3$, since their respective excess kurtosis have opposite signs---Eq.~\eqref{eq:sonine} tells us that the plotted ratio is basically $1+a_2 L_2^{1/2}(c^2)$. Point (ii)} allows us to check that $a_2^{\Frz}$---and thus the VDF---is indeed independent of $r_c$ for all inelasticities. {In addition, this} graph shows that $a_{2}^{\Frz}$ is really far from the steady-state kurtosis $a_2^{\st}$ but very close to the HCS values $a_2^{\hcs}$,  which suggests that the HCS has a key role in the frozen state---as further discussed in Appendix~\ref{sec:appendix-2}.

\subsection{Boundary layer approach. Universality}\label{sec:bl-universality}

We are now concerned with the behavior of the system for very low bath temperatures, when the system is close to its frozen state. 
% {Following the general theory for boundary layer problems~\cite{bender_advanced_1999}, a distinguished limit of the evolution equations \eqref{granularSonine} is found below by introducing suitably scaled variables for low bath temperatures.} In this way, we find an inner expansion, which is afterwards matched with the outer solution derived {before in Sec.~\ref{subsec:pert-mol}. Therefrom,} an (approximate) solution for all values of the bath temperature, known as a uniform solution~\cite{bender_advanced_1999}, is built.
To start with our boundary layer approach, we define the scaled variables
\begin{equation}\label{eq:scaling-cooling}
	Y \equiv  r_c^{-2/3}\theta, \quad X \equiv r_c^{-2/3}\theta_{\st} ,
\end{equation} 
as suggested by Eqs. \eqref{freezescaling} and \eqref{scaling-2}. Interestingly,  the evolution equations \eqref{granularSonine} become independent of the cooling rate when written in terms of $X$ and $Y$:
\begin{subequations}\label{eq:evolscaled}
\begin{align}
    -\frac{dY}{dX}&=  X^{3/2} \left(1+\frac{3}{16} a_2^{\st}\right) -Y^{3/2} \left(1+\frac{3}{16} a_2\right), \label{eq:evolscaled-Y} \\
    -\frac{da_2}{dX}&= 2\, Y^{1/2} \left[ \left(1-\frac{X^{3/2}}{Y^{3/2}}\right) a_2 +B \, (a_2^{\st} -a_2) \right]. 
\end{align}
\end{subequations}
These equations provide us with the inner solution, which is expected to be valid for $(X,Y,a_2)$ of the order of unity, i.e., close to the frozen state as discussed above. {Equations~\eqref{eq:evolscaled} are complemented} with the boundary conditions \eqref{eq:granular-Sonine-bc}, which {now read}
\begin{equation}\label{eq:boundary}
    Y(r_c^{-2/3}) = r_c^{-2/3}, \quad a_2(r_c^{-2/3}) = a_2^{\st}.
\end{equation}
Note that {all the dependence of the inner solution on the cooling rate $r_c$ takes place through} the boundary conditions.

\begin{figure}
	\centering
	\includegraphics[width=0.98\linewidth]{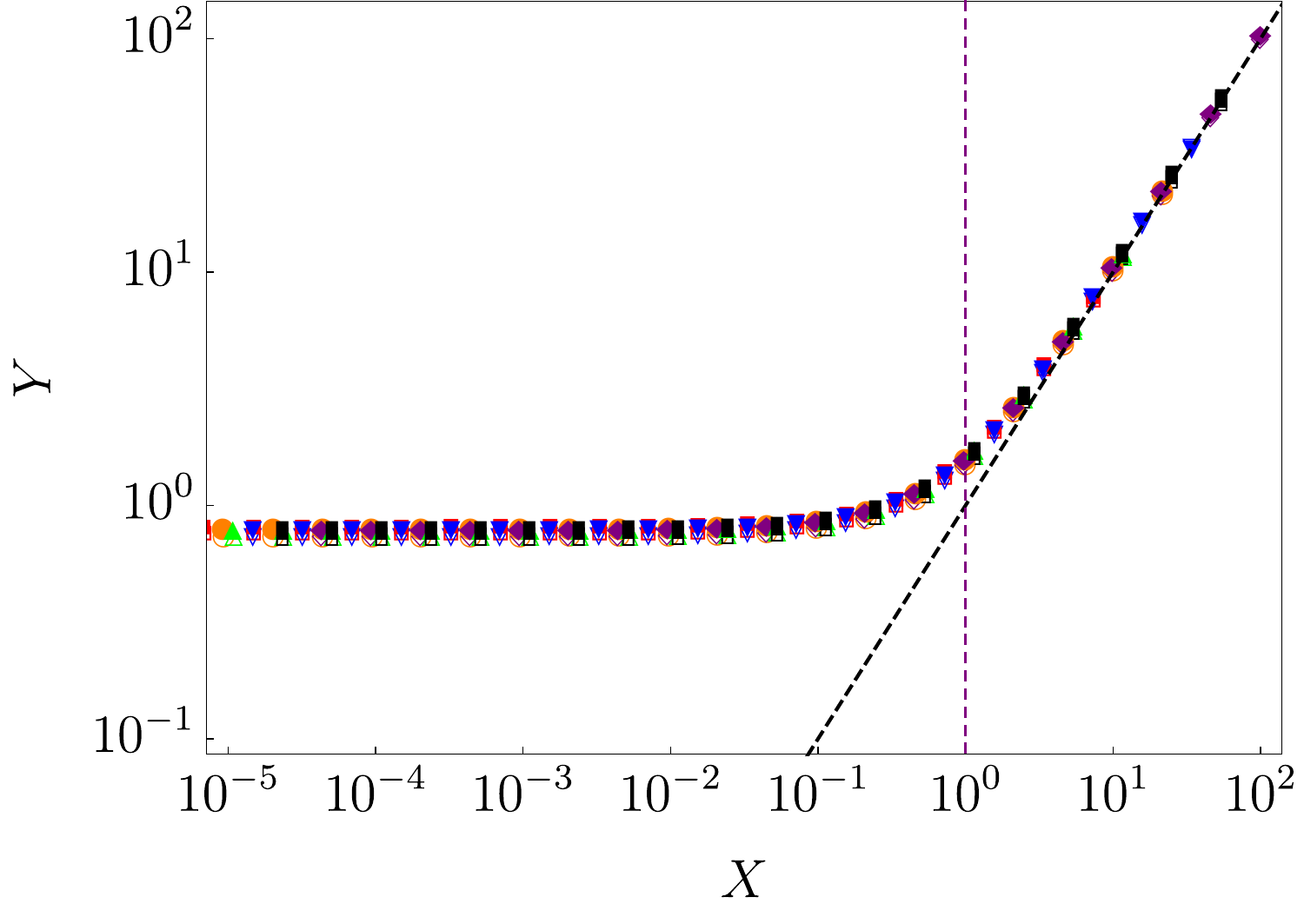} 
 \caption{{Scaled} granular temperature $Y$ as a function of {the scaled bath temperature} $X$. {We have employed the linear} cooling protocol \eqref{cooling-programme} {with} different cooling rates {and two values of the restitution coefficient $\alpha$}. The color codes and symbols for the DSMC data are the same as {those} {employed in the left panel of Fig.\ref{VDF-graph}. The dashed, purple vertical line marks the fictive temperature $X_f = \theta_f / r_c^{2/3} = 1$ from Eq.~\eqref{eq:thetaf-power-law}. for $c = 3/2$.}}
	\label{univ-cooling-gran}
\end{figure}

Figure~\ref{univ-cooling-gran} { illustrates the glass transition on the left panel of Fig.~\ref{frozen-temp-gran},} but in terms of the scaled variables $X$ and $Y$. { For each value of the restitution coefficient $\alpha$,} all the curves for different values of the cooling rate $r_c$ collapse onto a unique master curve, independent of $r_c$.  The only difference appears for large values of $X$, for which the different curves start from different initial points, consistently with the boundary conditions \eqref{eq:boundary}. { Our theoretical prediction for the scaled fictive temperature $X_f$ is also plotted: it is independent of $r_c$ as well, since $\theta_f$ is proportional to $r_c^{2/3}$, as given by Eq.~\eqref{eq:thetaf-power-law}, and $X_f=r_c^{-2/3} \theta_f$. Our theory thus gives an excellent estimate for the actual fictive temperature of the system.} Since the plotted numerical data corresponds to the DSMC integration of the kinetic equation \eqref{eq:BFPE}, not to our perturbation approach, this suggests that the exact solution to the problem presents a universal behavior in scaled variables.

{  The universal behavior illustrated in Fig.~\ref{univ-cooling-gran} depends mildly on $\alpha$; the differences between the $\alpha = 0.9$ (open symbols) and $\alpha = 0.3$ (filled symbols) datasets are very small. This is due to the smallness of the values of the excess kurtosis in the granular gas, where typically $|a_2|\lesssim 0.15$. Besides, note that the terms containing the excess kurtosis in the evolution equation for the temperature~\eqref{granularSonine-theta} are of the form $(1+3a_2/16)$, and thus $3a_2/16\lesssim 0.03$, so the differences between different values of $\alpha$ are expected to be of a few per cents. 
% This mild dependence is even more negligible for the fictive temperature $X_f$, which is approximately 1 up to the third digit.
}

{
In order to understand such universal behavior in scaled variables, it is useful to build approximate, to the lowest order, expressions over the whole bath temperature range, not only in the boundary layer. For the sake of having a clear, distinct notation, we denote such approximate expressions by $(Y^{\BL},a_{2}^{\BL})$. In Appendix~\ref{app:boundary-layer-granular}, we show that these lowest order expressions are given by the solution of Eq.~\eqref{eq:evolscaled} with the boundary conditions 
\begin{equation}
    \lim_{X\to\infty}Y^{\BL}(X) = \infty, \quad \lim_{X\to\infty} a_{2}^{\BL}(X) = a_2^{\st},
    \label{eq:boundary-infty}
\end{equation}
which are the limit as $r_c\to 0$ of Eq.~\eqref{eq:boundary}. Although it is not possible to write $(Y^{\BL}(X),a_{2}^{\BL}(X))$ in a simple closed form, it is clear that $(Y^{\BL}(X),a_{2}^{\BL}(X))$ does not depend on $r_c$, since neither the evolution equations~\eqref{eq:evolscaled} nor the boundary conditions~\eqref{eq:boundary-infty} depend on $r_c$. 
}

From the { lowest order} solution, the frozen values of the scaled variables are readily obtained,
\begin{subequations}
\label{eq:frozen-cond}
    \begin{align}
        Y^{\Frz}&\equiv \lim_{X\to 0}Y^{\BL}(X),
        \\
        a_2^{\Frz}&\equiv \lim_{X\to 0}a_{2}^{\BL}(X).
    \end{align}
\end{subequations}
Our above argument about the independence of $Y^{\BL}(X)$ on the cooling rate is immediately translated to $\theta_{\Frz}=r_c^{2/3}Y^{\Frz}$, which means that $\theta_{\Frz}$ follows the power law behavior $\theta_{\Frz}\propto r_c^{2/3}$ that we have already checked on the right panel of Fig.~\ref{frozen-temp-gran}. { Also, the independence of $a_2^{\Frz}$ on $r_c$ has already been checked on the right panel of Fig.~\ref{VDF-graph}.   Moreover, the independence on $r_c$ of the curves $(Y^{\BL},a_2^{\BL})$ as a function of $X$ gives rise to the universal cooling curve in Fig.~\ref{univ-cooling-gran}. Therefore, our theory explains the observed universal behavior of the simulation data in scaled variables.}

\section{Hysteresis cycles} \label{section4}

Now we turn our attention to a reheating protocol from the frozen state with rate $r_h$, $d\theta_{\st}/dt=+r_h$. First, we consider the paradigmatic case $r_h=r_c=r$. {Second, we consider the more general case $r_h \neq r_c$. In both cases, we show} that the system does not follow backwards the cooling curve, but crosses the NESS line $\theta=\theta_{\st}$ and afterwards tends thereto from below. This is similar to the hysteresis cycle displayed by glassy systems in temperature cycles (cooling followed by reheating).  

\subsection{Universal hysteresis cycle with $r_c = r_h$}

{First, we consider the case $r_c=r_h=r$. In complete analogy with the cooling program, we  define scaled variables as}
\begin{equation}\label{eq:scaling-hysteresis}
	Y \equiv r^{-2/3} \theta , \quad X \equiv r^{-2/3} \theta_{\st} .
\end{equation} 
In terms thereof, the evolution equations {in the reheating protocol become independent of $r$,}
\begin{subequations}\label{eq:evolscaled-heating}
\begin{align}
    \frac{dY}{dX}&=  X^{3/2} \left(1+\frac{3}{16} a_2^{\st}\right) -Y^{3/2} \left(1+\frac{3}{16} a_2\right),
    \label{eq:evolscaled-heating-temp}
    \\
    \frac{da_2}{dX}&= 2\, Y^{1/2} \left[  \left(1-\frac{X^{3/2}}{Y^{3/2}}\right) a_2 +B \, (a_2^{\st} -a_2) \right]. 
\end{align}
\end{subequations}
The above system must be complemented with the new boundary conditions
\begin{equation}\label{eq:evolscaled-heating-bc}
    Y(0) = Y^{\Frz}, \quad a_2(0) = a_2^{\Frz},
\end{equation}
which correspond to that of the frozen state from the previously applied cooling program, given by Eq.\eqref{eq:frozen-cond} to the lowest order---{recall that $r_c=r_h=r$.}

A completely similar analysis to that carried out for the cooling program shows that the solution to Eq.~\eqref{eq:evolscaled-heating}, i.e., the inner solution for the heating program, gives the uniform solution to the lowest order again. { The hysteresis cycle is unique in the rescaled axes $Y$ vs.~$X$, since the rate is nowhere in Eqs.~\eqref{eq:evolscaled-heating} and \eqref{eq:evolscaled-heating-bc}.}

\begin{figure*}
  \centering 
 {\includegraphics[width=3in]{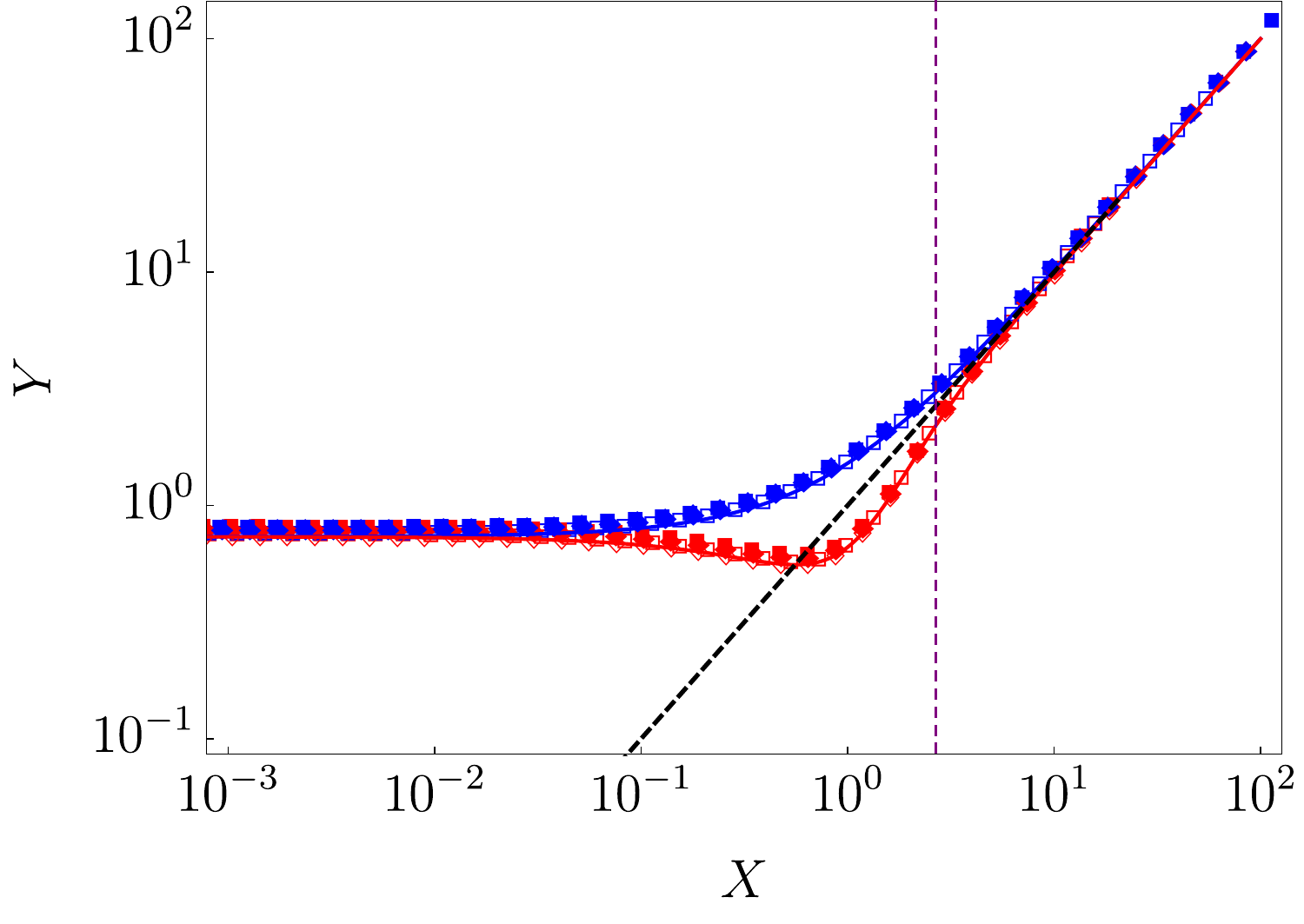} 
 \includegraphics[width=3in]{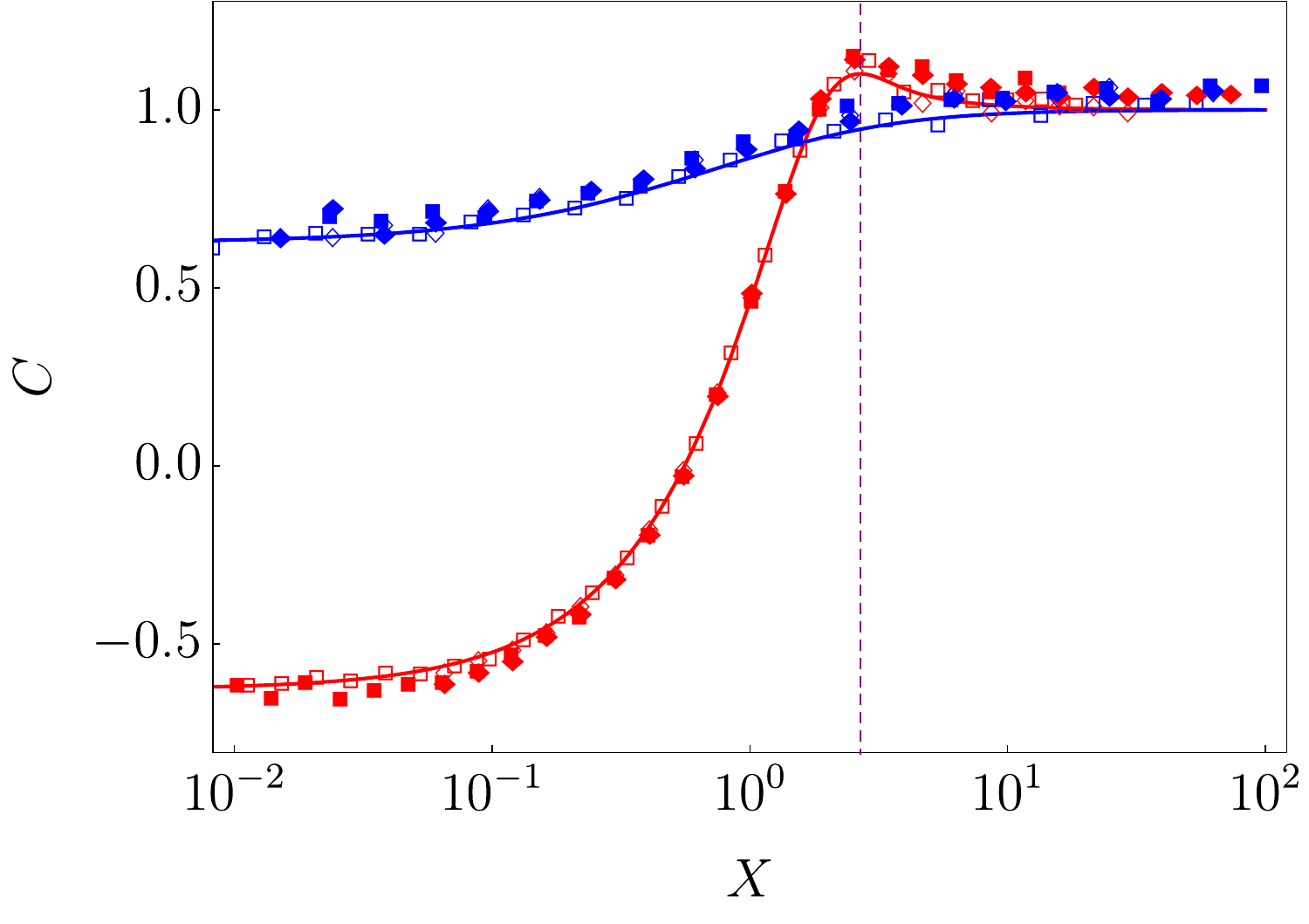}}
  \caption{{Hysteresis cycles in the granular gas. The system is first cooled down with rate $r_c$ and later reheated from the frozen state with rate $r_h=r_c=r$. Blue (red) symbols and lines correspond to the cooling (heating) protocol.} (Left) {Scaled} kinetic temperature {$Y=r^{-2/3}\theta$} as a function of the {scaled} bath temperature {$X=r^{-2/3}\theta_{\st}$. Specifically, we present results for $r=0.01$ (squares) and $r=0.001$ (diamonds), {and for two values of $\alpha$:  $0.9$ (open symbols) and $0.3$ (filled symbols).} } Symbols are simulation results of the {Boltzmann equation \eqref{eq:BFPE},} while the {solid} curves correspond to the numerical integration of Eqs.~\eqref{eq:evolscaled} and \eqref{eq:evolscaled-heating} {for cooling and heating, respectively.}  The dashed  straight line corresponds to the {instantaneous} NESS curve $Y =X$. {The purple vertical line marks the bath temperature $X_g$ at which the heat capacity reaches its maximum in the reheating program---see right panel. {Both the solid curves and the purple vertical line were obtained for $\alpha = 0.9$, as they superimpose with the ones corresponding to the $\alpha = 0.3$ case.}} (Right) Associated apparent heat capacity {$C=d\theta/d\theta_{\st}=dY/dX$. Again, the symbols have been obtained from the simulation data, and the lines correspond to the numerical integration of Eqs.~\eqref{eq:evolscaled} and \eqref{eq:evolscaled-heating}. Note the logarithmic scale used for the horizontal axis on both panels.}
}
  \label{hysteresis-gran}
\end{figure*}

{In Fig.~\ref{hysteresis-gran}, we numerically check our prediction on the hysteresis cycle being independent of $r$.} On the left panel, the hysteresis cycle of the kinetic temperature is shown. DSMC simulation data (symbols) are compared with the boundary layer solution (blue lines) of Eq.~\eqref{eq:evolscaled-heating}, for different values of the cooling/heating rate $r=r_c=r_h$, {and again for two values of the restitution coefficient $\alpha$: $0.3$ and $0.9$}. It is neatly observed that the boundary layer solution captures very well the numerical results throughout the whole cycle. Remarkably, the heating curve crosses the NESS line $\theta=\theta_{\st}$ (dashed line) and tends thereto from below---this is further analyzed in Sec.~\ref{sec:normal-solution}. On the right panel, we display the apparent ``heat capacity" $d\theta/d\theta_{\st}=dY/dX$ over the thermal cycle. {In general, the apparent} heat capacity
\begin{equation}
    C \equiv \frac{d\theta}{d\theta_{\st}}=\frac{dY}{dX}
\end{equation}
is  nonmonotonic in the heating process, with a marked maximum at a certain value of $\theta_{\st}$ (or $X$) that can be employed to define the glass transition temperature $\theta_{g}$ (or $X_g$)~\cite{brey_dynamical_1994,angell_formation_1995,dyre_colloquium_2006,tropin_modern_2016,richet_thermodynamics_2021}. In this simple system, the value of the apparent heat capacity  has opposite signs in the limit of very low bath temperatures. This is readily understood from Eqs.~\eqref{eq:evolscaled-Y} and \eqref{eq:evolscaled-heating-temp}, since
\begin{equation}\label{eq:C-low-temp-gran}
    C\sim \pm Y^{3/2}\left(1+\frac{3}{16}a_2\right), \quad X\ll 1,
\end{equation}
with the plus and minus signs corresponding to cooling and reheating, respectively.

{
The behavior of the apparent heat capacity $C$ in Eq.~\eqref{eq:C-low-temp-gran} has a neat physical meaning. For very low bath temperatures, the term coming from the stochastic driving in the evolution equation for the temperature~\eqref{granularSonine-theta} becomes negligible and
\begin{equation}
\frac{d\theta}{dt}\sim -\theta^{3/2}\left(1+\frac{3}{16}a_2\right).
\label{eq:theta-low-temp}
\end{equation}
That is, the granular gas ``freely cools'', since $\theta$ mononotically decreases with time. Note that, in fact, Eq.~\eqref{eq:theta-low-temp} is nothing but Haff's law, $\dot\theta\propto -\theta^{3/2}$, and it is equivalent to  Eq.~\eqref{eq:C-low-temp-gran} for the apparent heat capacity.
}

\subsection{Normal heating curve}\label{sec:normal-solution}

In order to {deepen our understanding of the hysteretic behavior in reheating, a regular perturbation expansion can be carried out---analogous to the one for the cooling process. By simply substituting $r_c$ with $-r_h$ in Eq.~\eqref{outer-gran-cooling}, we obtain}
\begin{subequations}
\label{outer-gran}
		\begin{align}
			\theta &= \theta_{\st}- \frac{2\,r_h}{3\theta_{\st}^{1/2}}  \left[1+\frac{3}{16} \; a_2^{\st} \left( 1+\frac{1}{B} \right) \right]^{-1}+O(r_h^2),  \label{outer_granular}
			\\
			a_{2} &= a_2^{\st}+\frac{r_h \,a_2^{\st}}{\,B\,\theta_{\st}^{3/2}}  \left[1+\frac{3}{16} \; a_2^{\st} \left( 1+\frac{1}{B} \right) \right]^{-1}+O(r_h^2). 
		\end{align}
	\end{subequations}
These perturbative expressions are expected to be valid for not too low temperatures {$\theta_{\st}\gg r_h^{2/3}$,} i.e., over the outer layer---{using once more} the terminology of boundary layer theory. 

{Equations~\eqref{outer-gran}} depend on the heating program $r_h$, but not on the previously applied cooling program with cooling rate $r_c$. In other words, if we start the heating process from different initial frozen temperatures $\theta^{\Frz} = Y^{\Frz} r_c^{2/3}$ corresponding to different values of $r_c$ but reheat with a common rate $r_h$, we expect to approach the behavior in Eq.~\eqref{outer-gran} once the system reaches the outer layer.

\begin{figure}
	\centering
	\includegraphics[width=0.98\linewidth]{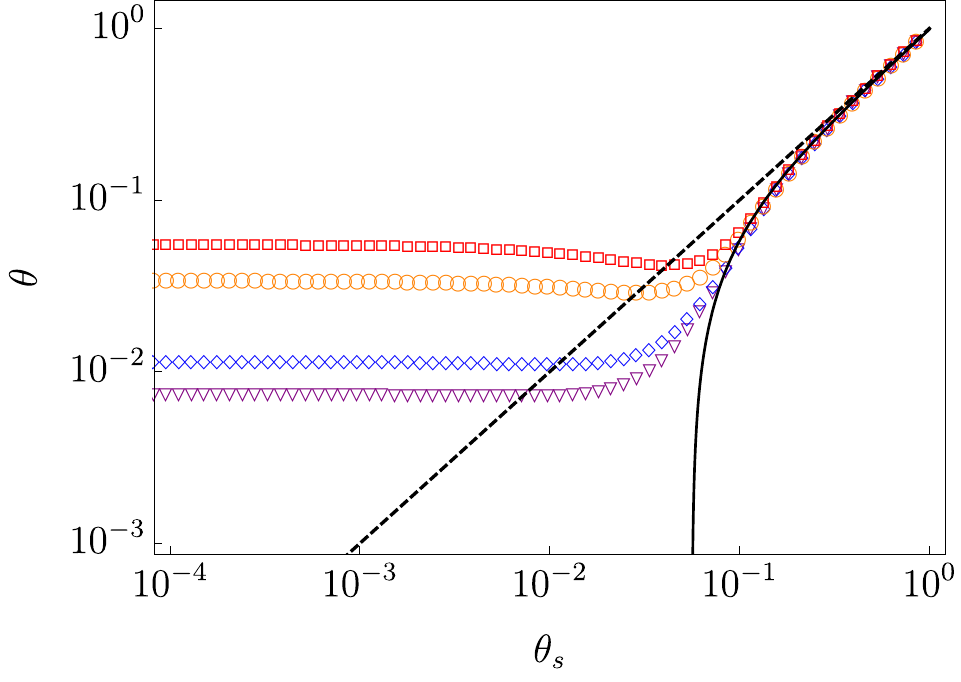} 
 \caption{Hysteresis cycles for reheating with rate $r_h$ from the frozen states corresponding to different cooling rates $r_c$. {All} reheating curves correspond to $r_h = 0.01$, and the different cooling rates employed are: (color, $r_c$) = (red, $0.05$), (orange, $0.01$), (blue, $0.005$) and (purple, $0.001$). Symbols correspond to DSMC simulation data, whereas the  solid {line} corresponds to the perturbative expression for the normal curve in Eq.~\eqref{outer-gran}.}
	\label{univ-heating} 
\end{figure}

The behavior just described above is illustrated in Fig.~\ref{univ-heating}: despite having different cooling programs, all the reheating curves tend towards a universal curve, independent of $r_c$, for high enough values of the bath temperature. Equation~\eqref{outer_granular} explains why the kinetic temperature overshoots the NESS curve $\theta=\theta_{\st}$ in reheating. The universal curve for the temperature, as given by Eq.~\eqref{outer_granular} in the outer layer, is always below the NESS curve---whereas the cooling curves always lie above the NESS curve, as expressed by Eq.~\eqref{outer-gran-cooling-temp} and illustrated by Fig.~\ref{frozen-temp-gran}. 

{From a physical standpoint, the overshoot of the instantaneous NESS curve may be understood by taking into account that the kinetic temperature $\theta$ always lags behind the bath temperature $\theta_{\st}$ during the entire time evolution of the hysteresis cycle. In the cooling protocol, this implies that the deviation $\theta- \theta_{\st}$ increases as $\theta_{\st}$ decreases, i.e. as the characteristic relaxation time of $\theta$ increases. In the reheating protocol, $\theta$ initially decreases, specifically as long as $\theta_{\st} \ll \theta$. It is not until $\theta_{\st} \simeq \theta$ that $\theta$ starts to increase but, at the time it does, as its characteristic relaxation time is still large compared to the reheating time, we will have $\theta_{\st}>  \theta$ until it reaches the instantaneous NESS. The latter argument also explains why the crossing points between the NESS curve and the simulation data from Fig.~\ref{univ-heating} are very close to the minimum ---i.e. the point at which $d\theta/d\theta_{\st}=0$--- of each dataset.}

{Let us have a more detailed look at the dependence of the reheating curves on the rate $r_c$ of the previous cooling program. In Fig.~\ref{univ-heating}, it is observed that the reheating curves develop a neat dimple as the cooling rate $r_c$ is increased. This can be understood by going back to the evolution equations for the inner region \eqref{eq:evolscaled-heating}, which continue to be valid for $r_c\ne r_h$, but with the new boundary conditions 
\begin{equation}\label{eq:bc-reheating-rc-ne-rh}
    Y(0) = (r_c/r_h)^{2/3} Y^{\Frz}, \quad a_2(0) = a_2^{\Frz}.
\end{equation}
The boundary conditions in Eq.~\eqref{eq:evolscaled-heating-bc} can be thus considered as the particularization of Eq.~\eqref{eq:bc-reheating-rc-ne-rh} to the case $r_h=r_c$. %Equation~\eqref{eq:evolscaled-heating-temp} predicts that
% \begin{equation}\label{eq:dimple}
%     \frac{d\theta}{d\theta_{\st}}=\frac{dY}{dX}\sim -Y^{3/2}\left(1+\frac{3}{16}a_2\right)<0, \quad X\ll 1.
% \end{equation}
%   THIS WAS NEARLY IDENTICAL TO THE EQUATION FOR $C$ JUST ABOVE
As $r_c/r_h$ increases, we have that $Y(0)=(r_c/r_h)^{2/3} Y^{\Frz}$ also increases and then the initial decrease { predicted by Eq.~\eqref{eq:C-low-temp-gran} for the heating curve} becomes more noticeable, giving rise to the neat minimum (the dimple) shown by the uppermost curve in Fig.~\ref{univ-heating}. For small values of $r_c/r_h$, this initial decrease is barely noticeable, and the tendency towards the normal curve is almost horizontal, as shown by the lowermost curve in Fig.~\ref{univ-heating}. Still, it must be noted that $\theta$ always presents a minimum as a function of $\theta_{\st}$, because ${d\theta}/{d\theta_{\st}}$ is negative for very low temperatures, as predicted by { Eq.~\eqref{eq:C-low-temp-gran} for heating,} whereas it is positive, since ${d\theta}/{d\theta_{\st}}\to 1$, for high temperatures.} 

% {It is also worth mentioning that the behavior towards such universal curve varies depending on the previously applied cooling program: for rapid coolings (i.e high values of $r_c$), the corresponding heating curves present more pronounced dimples than for slow coolings. In fact, for $r_c = 0.001$, the heating curve is almost monotonous, which reflects the fact that the nonmonotonic behavior of the heat capacity depends on the ratio between the heating and cooling rates.}

The {approach to a unique curve, independent of the previous cooling program, of the granular gas upon  reheating  is similar to the behavior found in models} described by master equations. {Therein, it has been} analytically proved that there exists a universal \textit{normal} curve that is the global attractor of the dynamics for heating processes~\cite{brey_normal_1993,brey_dynamical_1994,brey_dynamical_1994-1,prados_hysteresis_2000,prados_glasslike_2001}. The expressions in Eq.~\eqref{outer-gran} may be thus regarded as the regular perturbation expansions of a similar normal curve in the granular gas.

\begin{figure}
	\centering
\includegraphics[width=0.98\linewidth]{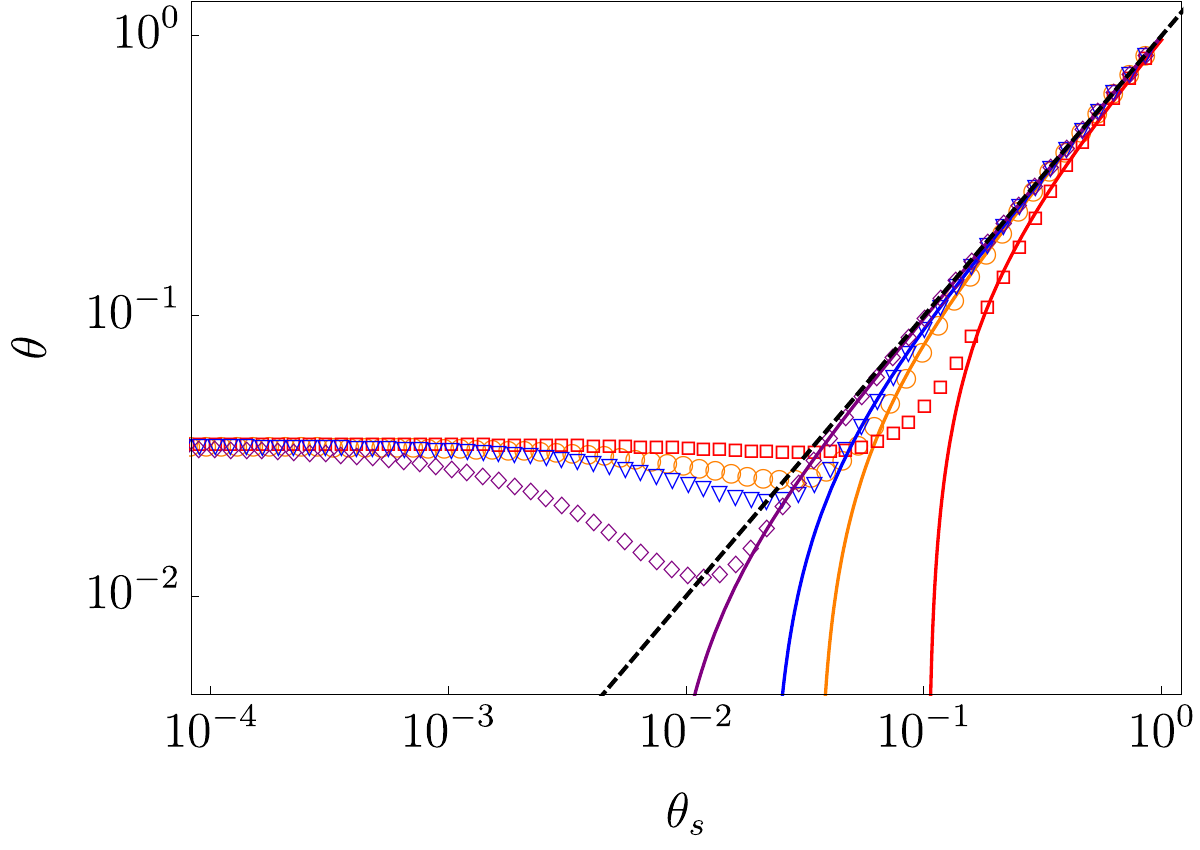} 
 \caption{{Hysteresis cycles for reheating with different rates $r_h$ from the common frozen state corresponding to a given value of $r_c$. Specifically, the plotted data correspond to $r_c=0.01$ and the reheating rates:  (color, $r_h$) = (red, $0.05$), (orange, $0.1$), (blue, $0.005$) and (purple, $0.001$). As in Fig.~\ref{univ-heating}, symbols correspond to DSMC simulation data, whereas the solid curves correspond to Eq.~\eqref{outer-gran}.}}
	\label{further-normal-curve}
\end{figure}

{The tendency towards the normal curve is further illustrated in Fig.~\ref{further-normal-curve}, in which we consider reheating with different values of $r_h$ from a common frozen state, corresponding to one value of $r_c$. For all the values of $r_h$, the hysteresis cycles cross the instantaneous NESS curve $\theta=\theta_{\st}$, since the corresponding normal curves always lie below it. Moreover, the hysteresis cycle is larger as $r_h$ increases, because the normal curve is more distant from the NESS curve. Note that the heating curve develops a neat dimple as $r_h$ decreases---consistently with our discussion above, which told us that the minimum gets more marked as $r_c/r_h$ increases.}

\begin{figure}
	\centering	\includegraphics[width=0.95\linewidth]{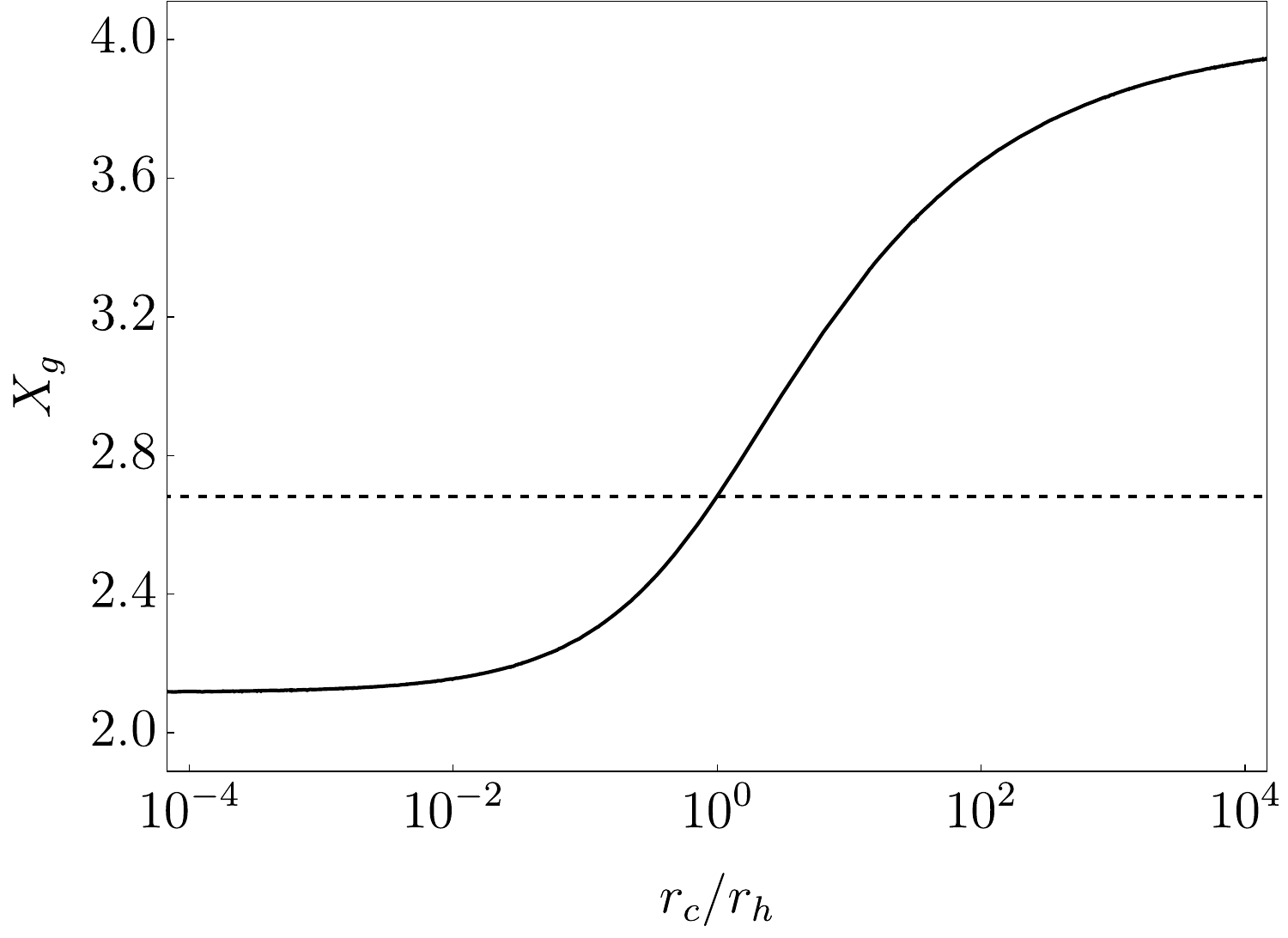} 
 \caption{{Scaled glass transition temperature $X_g=r_h^{-2/3}\theta_g$ for the granular gas as a function of the ratio $r_c/r_h$. Only plotted is the theoretical prediction for $X_g$, obtained via the numerical integration of the evolution equations for the scaled variables, in a hysteresis cycle characterized by a cooling with rate $r_c$ followed by reheating with rate $r_h$. The dashed horizontal line marks the value for $r_c = r_h$---which corresponds to  the vertical line on both panels of Fig.~\ref{hysteresis-gran}.}}
	\label{transition-temperature}
\end{figure}

{The crossing of the instantaneous stationary curve $\theta=\theta_{\st}$ stemming from the tendency towards the normal curve entails that the apparent heat capacity $C$ upon reheating always displays a maximum at a certain bath temperature. The position of this maximum can be employed to define a glass transition temperature $\theta_{g}$---or a scaled one $X_g=r_h^{-2/3}\theta_g$. Our theory predicts that $X_g$ only depends on the ratio $r_c/r_h$, since the evolution in scaled variables in reheating is governed by Eqs.~\eqref{eq:evolscaled-heating} with the boundary conditions in Eq.~\eqref{eq:bc-reheating-rc-ne-rh}. Only the latter introduce dependence on the rates; namely on their ratio $r_c/r_h$. In Fig.~\ref{transition-temperature}, we plot this theoretical prediction for $X_g$ as a function of the ratio $r_c/r_h$. Note that $X_g = O(1)$ regardless of the value of the ratio $r_c/r_h$, which entails that the glass transition temperature $\theta_g=r_h^{2/3} X_g$ is basically proportional to $r_h^{2/3}$ in the granular gas.} { We also highlight that $X_g$ is of the same order as the fictive temperature $X_f$ from Eq.~\eqref{eq:thetaf-power-law}, which is consistent, as both temperatures give a qualitative account of the glass transition.}

\section{Molecular fluid with nonlinear drag} \label{section5}

We now focus our attention {on} a second relevant physical system: a molecular fluid with nonlinear drag~\cite{klimontovich_statistical_1995,lindner_diffusion_2007,santos_mpemba_2020,patron_strong_2021,goychuk_nonequilibrium_2021}. The considered model arises when analyzing an ensemble of Brownian particles of mass $m$ immersed in an isotropic and uniform background fluid~\cite{ferrari_particles_2007,ferrari_particles_2014}, the particles of which have mass $m_{\text{bf}}$. In the $\mbf/m\to 0$ limit---the so-called Rayleigh limit, the drag coefficient $\zeta$ becomes velocity independent and thus the drag force is linear. However, in real physical scenarios we have that $\mbf/m\ne 0$, and it is thus relevant to consider the corrections to the Rayleigh limit. Specifically, by introducing the first order corrections thereto, i.e., by retaining only linear terms in $\mbf/m$, the drag coefficient is found to be quadratic on the velocities~\cite{ferrari_particles_2007,ferrari_particles_2014,hohmann_individual_2017}. Interestingly, it has recently been shown that this model {accurately} describes a mixture of ultracold Cs and Rb atoms~\cite{hohmann_individual_2017}.

Let us consider a system of $d$-dimensional hard spheres of mass $m$, diameter $\sigma$, and density $n$ immersed in a background fluid at temperature $T_{\st}$. In the regime just explained above, the Brownian particles are subjected to a nonlinear drag force of the form~\cite{ferrari_particles_2007,ferrari_particles_2014,hohmann_individual_2017}
\begin{equation}
	\bm{F}=-m \, \zeta(v) \bm{v}, \label{drag} 
\end{equation}
where $\bm{v}$ is the particle velocity, {and}
\begin{equation}
\label{drag-coeff}
	\zeta(v)=\zeta_0 \Big( 1+\gamma\frac{m v^2}{k_B T_{\st}}\Big) 
\end{equation}
is a nonlinear drag coefficient. {Therein, $\zeta_0$ is its zero-velocity limit and the dimensionless parameter $\gamma$ measures the
degree of nonlinearity. For} hard spheres, it is found that $\zeta_0 \propto T_{\st}^{1/2}$---see e.g., Refs.~\onlinecite{hohmann_individual_2017,santos_mpemba_2020} for the complete expression. The dependence of $\zeta_0$ on $T_{\st}$ is relevant here because the bath temperature depends on time in cooling/heating processes.

Similarly to the granular gas, the system may be accurately described by the one-particle VDF $f(\bm{v},t)$ if sufficiently dilute. In this case, the dynamical evolution of the VDF is governed by the Fokker-Planck equation (FPE) 
\begin{equation}
	\label{FPE}
	\partial_t f(\bm{v},t) - \frac{\partial}{\partial \bm{v}}\cdot
	\left[ \zeta(v) \bm{v} +\frac{\xi^2(v)}{2}\frac{\partial}{\partial
		\bm{v}}\right ]f(\bm{v},t)  = 0,
\end{equation}  
where $m^2 \xi^2(v)$  is the variance of a stochastic white noise force.  The coefficients $\xi^2 (v)$ and $\zeta (v)$ are  related by means of the fluctuation-dissipation relation 
\begin{equation}
	\xi^2(v) = \frac{2k_BT_{s}}{m} \zeta(v),
\end{equation}
which ensures that the equilibrium Maxwellian VDF 
\begin{equation}
	\label{eqfdv}
	f_s(\bm{v}) = n \left( \frac{m}{2\pi k_B T_{s}} \right)^{\frac{d}{2}} e^{-\frac{mv^2}{2k_BT_{s}}}, 
\end{equation}
constitutes the unique stationary solution of the FPE~\eqref{FPE}. 

The velocity dependence of the drag coefficient implies that we have multiplicative noise in this problem~\cite{van_kampen_stochastic_1992,gardiner_stochastic_2009}. By employing the Ito interpretation of stochastic integration~\cite{van_kampen_ito_1981,mannella_ito_2012}, which is the most convenient one for numerical simulations, the FPE is equivalent to the following Langevin equation: 
\begin{equation}
	\label{langevin}
	\dot{\bm{v}}(t) = -\zeta_{\rm{eff}}(v) \, \bm{v}(t) + \xi(v) \, \bm{\eta}(t),  
\end{equation}
where 
\begin{equation}
	\zeta_{\rm{eff}}(v)=\zeta_0\Big( 1-2 \gamma +\gamma\frac{m v^2}{k_B T_{\st}}\Big) 
\end{equation}
constitutes an effective drag coefficient, while $\bm{\eta}(t)$ is a Gaussian
white noise of zero average $ \langle \bm{\eta}(t) \rangle=0$ and
correlations $ \langle \eta_i(t) \, \eta_j(t') \rangle=\delta_{i,j} \,
\delta (t-t')$.

{The kinetic temperature is again defined as in Eq.~\eqref{suppl-temperature-gran} for the granular gas, but understanding $f(\bm{v},t)$ as the solution of the FPE.} Inserting \eqref{suppl-temperature-gran} into \eqref{FPE} leads to the following evolution equation for the temperature,
\begin{align}
    \dot{T} = \zeta_0 \; \biggl\{ & 2(T_{\st}-T)\left[1+\gamma (d+2)\frac{T}{T_{\st}}\right]  
    - 2\gamma(d+2)\frac{T^2}{T_{\st}}a_2 \biggr\},  \label{suppl-evol-eq-temp-dim}
\end{align}
where $a_2$ corresponds to the excess kurtosis, previously introduced in Eq.~\eqref{suppl-excess-kurtosis} when studying the granular gas. 

For nonlinear drag, $\gamma \ne 0$, the evolution of the
temperature is coupled to that of the excess kurtosis and, thus, we
need to consider the evolution equation for the latter too. In turn,
the evolution equation for the excess kurtosis involves sixth-degree
moments, and in general there emerge an infinite hierarchy of equations
for the moments. 
Under the first Sonine approximation, we have the evolution equations~\cite{santos_mpemba_2020,patron_strong_2021}
\begin{subequations}\label{suppl-eq:evol-eqs-first}
\begin{align}
    \dot{\theta}=   \theta_{\st}^{1/2} \, \bigg[ & 2(\theta_{\st}-\theta) + 2\gamma (d+2)\, \theta 
    % \nonumber \\ 
    % & 
    - 2\, \gamma \, (d+2)(1+a_2)\frac{\theta^2}{\theta_{\st}} \bigg],\label{suppl-eq:T-evol}\\
    \dot{a}_2=  \theta_{\st}^{1/2} \; \biggl\{&  8\gamma \left(1-\frac{\theta}{\theta_{\st}}\right) 
    % \nonumber \\
    %  & 
     -\left[\frac{4\theta_{\st}}{\theta}-8\gamma+4\gamma(d+8)\frac{\theta}{\theta_{\st}} \right]a_2\biggr\}, \label{suppl-eq:a2-evol} 
\end{align}
\end{subequations}
where we have introduced the dimensionless variables
\begin{equation}
	\label{suppl-dimensionless-variables-1}
	\theta \equiv
	\frac{T}{T_i}  , \quad \theta_{\st} \equiv \frac{T_{\st}}{T_i}, \quad 
    t^*\equiv \zeta_{0}(T_i) \; t,
\end{equation}
with $T_i \equiv T(t=0)$ being the initial temperature. We have also taken into account that $\zeta_0(T_{\st})=\zeta_0(T_i)\theta_{\st}^{1/2}$.

In previous work~\cite{patron_strong_2021,patron_nonequilibrium_2023}, we have shown that the nonlinear fluid approaches  a nonequilibrium state, termed LLNES (long-lived nonequilibrium state), over a wide intermediate timescale, when instantaneously quenched to low enough values of the bath temperature, i.e., $T_i/T_{\st} \gg 1$. The VDF at the LLNES is given by a delta peak; in terms of the scaled variables in Eq.~\eqref{eq:scaled-variables}, it reads
\begin{equation}
    \label{eq:llnes-sol}
    \phi_{\LLNES}(\bm{c}) = \Omega_d^{-1} \left(\frac{2}{d}\right)^{\frac{d-1}{2}} \delta \!\left(c - \sqrt{\frac{d}{2}}\right),
\end{equation}
with $\Omega_d$ being the $d$-dimensional solid angle~\cite{patron_nonequilibrium_2023}. The exact value of the excess kurtosis at the LLNES will be useful, which is
\begin{equation}\label{eq:a2-L}
    a_2^{\LLNES}=-\frac{2}{d+2}.
\end{equation}
It is worth remarking that the VDF for the LLNES, and thus $a_2^{\LLNES}$, does not depend on the nonlinearity parameter $\gamma$~\cite{patron_nonequilibrium_2023}.

The LLNES state corresponds to the extreme scenario that comes about when the system is instantaneously quenched to a very low  temperature. In this case, for a system relaxing from equilibrium at $T_i$ to equilibrium at $T_{\st}\ll T_i$, the system first reaches the LLNES and afterwards tends to equilibrium from it. Note the strong similarity with the HCS for granular gases, which also appears when the intensity of the stochastic thermostat is instantaneously quenched to a very low value. In such a protocol, the granular gas first approaches the HCS and afterwards tends to the stationary state imposed by the stochastic thermostat. Thus, it is worth investigating the role played by the LLNES in the possible emergence of a {kinetic} glass transition in fluids with nonlinear drag.

\section{Glassy behavior of the nonlinear molecular fluid}\label{section-glass-molecular}

Now, in order to investigate a possible glass transition in the molecular fluid, we decrease the bath temperature following the same cooling program as in Eq.~\eqref{cooling-programme} for the granular gas.
{ The physical reason for the emergence of a kinetic glass transition is completely similar to that for the granular gas: in the nonlinear molecular fluid, the characteristic timescale for relaxation is determined by $\zeta_0^{-1}$, which also diverges as $T_{\st}^{-1/2}$ for low bath temperatures. Therefore, we expect the same scalings with the cooling rate as in the granular gas---derived by physical arguments from this divergence in Sec.~\ref{sec:why-kinetic-glass-transition}.}

In fact, as we follow the same perturbative approach in the cooling rate, we leave the mathematical details for Appendix \ref{app:reg-pert-mol-fluid}. Up to order $O(r_c)$, the regular perturbation solution {reads}
 \begin{subequations}
 \label{eq:perturbative-gran-o1}
     \begin{align}
         \theta &=\theta_{\st} +  \frac{r_c}{2\theta_{\st}^{1/2}}\, \frac{1+\gamma (d+6)}{\left[1 + \gamma (d+4)\right]^2 - 2\gamma^2
   (d+4)},
         \\
         a_2 &= -\frac{r_c}{\theta_{\st}^{3/2}}\frac{\gamma}{\left[1 + \gamma (d+4)\right]^2 - 2\gamma^2 (d+4)}.
     \end{align}
 \end{subequations}

Thus, we have $\theta-\theta_{\st} \propto r_c/ \theta_{\st}^{1/2}$ and $a_2\propto r_c/ \theta_{\st}^{3/2}$. Our regular perturbative approach fails when the $O(r_c^0)$ and  the $O(r_c^1)$ terms become comparable, i.e., again when $\theta_{\st} = O(r_c^{2/3})$, which implies that $\theta=O(r_c^{2/3})$ and $a_2=O(1)$. Let us {remark} that, regardless of the intrinsic differences between the molecular fluid and granular gas systems, they lead to the same scaling for both the kinetic temperature and the excess kurtosis.

The above discussion entails the necessity of introducing again a boundary layer approach. We define scaled variables, analogous to those for the granular gas in Eq.~\eqref{eq:scaling-cooling}, 
$Y \equiv  r_c^{-2/3} \theta$ and  $X \equiv r_c^{-2/3} \theta_{\st}$. In term of the scaled variables, the evolution equations~\eqref{suppl-eq:evol-eqs-first} become independent of $r_c$,
\begin{subequations}
\label{ODESinnerc}
\begin{align}
-\frac{dY}{dX} =  X^{1/2} \, \bigg\{ & 2\, ( X-Y)\left[1+\gamma (d+2)\frac{Y}{X}\right] 
\nonumber \\ & 
-2\gamma(d+2)\frac{Y^2}{X} a_2 \bigg\},  \label{ODESinnerc-Y}\\
	-\frac{da_2}{dX} = X^{1/2}\bigg\{ & 8\gamma \left(1-\frac{Y}{X}\right)  \nonumber
    \\
    &-\left[\frac{4X}{Y}-8\gamma +4\gamma (d+8)\frac{Y}{X} \right]a_2  \bigg\}.
\end{align}
\end{subequations}
{It is the boundary conditions that absorb all the dependence on $r_c$,}
\begin{equation}\label{eq:boundary-molecular}
    Y(r_c^{-2/3}) = r_c^{-2/3}, \quad a_2(r_c^{-2/3}) = 0.
\end{equation}

{The resemblance between the above framework and that of our previous study for the granular gas is neat. To avoid reiteration, we thus} focus on the main aspects of the glassy behavior in the molecular fluid. As will be seen, the analogy with the behavior found in the granular gas is almost complete. 

The lowest order solution for the cooling protocol would be again obtained by solving Eqs.~\eqref{ODESinnerc} with the boundary conditions $\lim_{X\to\infty}Y(X) =\infty$, $\lim_{X\to\infty}a_2(X) = 0$, which is completely independent of $r_c$.  At the frozen state we thus have
\begin{align} \label{eq:frozen-nonlinear}
    Y^{\Frz}\equiv \lim_{X\to 0}Y(X), \quad 
        a_2^{\Frz}\equiv \lim_{X\to 0}a_{2}(X),
\end{align}
{which are independent of $r_c$. We check this theoretical prediction with numerical data} in Table~\ref{table1}, in which we compare the value of $Y^{\Frz}$ and $a_2^{\Frz}$ obtained from numerical simulation of the Langevin equation~\eqref{langevin} and our theoretical prediction~\footnote{That is, the numerical integration of Eq.~\eqref{ODESinnerc} with the boundary conditions $\lim_{X\to\infty}Y(X)=\infty$,  $\lim_{X\to\infty}a_2(X)=0$.} {for different values of $r_c$}. The agreement is excellent for the kinetic temperature, and fair for the excess kurtosis. This was to be expected within the first Sonine approximation, since $a_2^{\Frz}$ is quite large for the nonlinear fluid. Moreover, $a_2^{\Frz}$ is not {as} close to its value  at the LLNES, $a_2^{\text{L}}=-0.4$ for $d=3$ as predicted by Eq.~\eqref{eq:a2-L}, as it was $a_2^{\Frz}$ close to its HCS value in the granular gas. See Appendix~\ref{sec:appendix-2} for a more detailed discussion on this point.
\begin{table}
\centering
 \begin{tabular}{|| P{2.5 cm} | P{1. cm} | P{1. cm}  | P{1. cm}  |} 
 \hline
   & $Y^{\Frz}$ & $a_2^{\Frz}$ \\ [0.5ex] 
 \hline
  \text{Boundary layer} & 0.397&-0.154  \\ [1ex]
 \hline
 \text{Sim.} ($r_c = 0.05$) & 0.402 & -0.146 \\ [1ex]
 \hline
 \text{Sim.} ($r_c = 0.01$) & 0.403 & -0.147 \\ [1ex]
 \hline
 \text{Sim.} ($r_c = 0.005$) & 0.403 & -0.144\\ [1ex]
 \hline
 \text{Sim.} ($r_c = 0.001$) & 0.404 & -0.148 \\ [1ex]
 \hline
\end{tabular}
\caption{Comparison between the numerical (simulation) and theoretical (boundary layer) values of the scaled kinetic temperature and the excess kurtosis at the frozen state. Specifically, we have considered a {molecular} fluid with nonlinearity parameter $\gamma = 0.1$ {in the three-dimensional case $d=3$. The same values of the parameters are employed in the remainder of the numerical simulations for the nonlinear fluid presented in this work.}}
\label{table1}
\end{table}

\begin{figure}
  \centering 
 \includegraphics[width=0.98\linewidth]{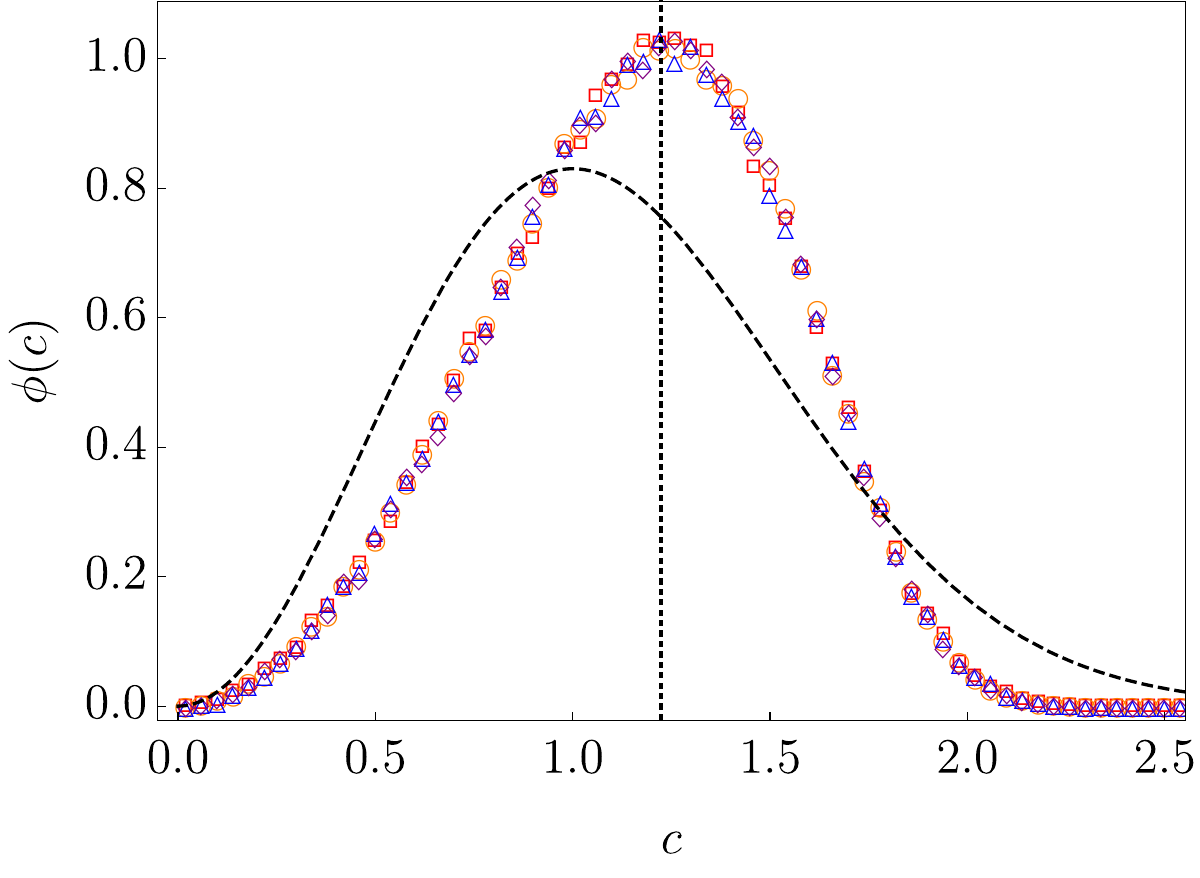}
  \caption{Plot of the dimensionless VDF at the frozen state for the nonlinear fluid. Symbols correspond to the numerical integration of the Langevin equation {with} $N= 10^5$ stochastic trajectories for different cooling rates: $r_c= 0.005$ (purple diamonds), $0.01$ (blue triangles), $0.05$ (orange circles) and $0.1$ (red squares).  The dashed curve corresponds to the equilibrium Maxwellian, whereas the dotted line marks the position of the {LLNES} Dirac-delta peak, as given by Eq.~\eqref{eq:llnes-sol}. 
  %Other parameters are $\gamma = 0.1$, $d=3$.
  }
  \label{frozen-temp-molec}
\end{figure}

The independence of $a_2^{\Frz}$ on the cooling rate suggests that this property should also hold for the complete VDF of the nonlinear fluid---as was the case for the granular gas. We check this property {in Fig.~\ref{frozen-temp-molec},} by plotting the scaled VDF for the nonlinear fluid in the frozen state, obtained from the numerical integration of the Langevin equation~\eqref{langevin}. The universality of the VDF at the frozen state is clearly observed. The largeness of $a_2^{\Frz}$ entails that the deviation from the Maxwellian equilibrium distribution is also large. For reference, the position of the delta peak corresponding to the LLNES is also plotted.

From the frozen state, we may reheat the system with the same rate $r_h=r_c$. Once more, scaled variables are introduced as $Y \equiv r_h^{-2/3} \theta$ , $X \equiv r_h^{-2/3} \theta_{\st}$, and the evolution equations become independent of the heating rate
\begin{subequations}
\label{ODESinnerh}
\begin{align}
\frac{dY}{dX} =  X^{1/2} \, \bigg\{  & 2\, ( X-Y)\left[1+\gamma (d+2)\frac{Y}{X}\right]
%\nonumber \\
%    & 
    -2\gamma(d+2)\frac{Y^2}{X} a_2 \bigg\},  \\
	\frac{da_2}{dX} = X^{1/2}\bigg\{ & 8\gamma \left(1-\frac{Y}{X}\right)  
 % \nonumber
%    \\
%    & 
    -\left[\frac{4X}{Y}-8\gamma +4\gamma (d+8)\frac{Y}{X} \right]a_2  \bigg\}.
\end{align}
\end{subequations}
{Note that the heating evolution equations} differ from the cooling evolution equations \eqref{ODESinnerc} only in the sign of the left hand side (lhs). Again, this system has to be solved with the boundary conditions $Y(0) = Y^{\Frz}$,  $a_2(0) = a_2^{\Frz}$, which correspond to the frozen state from the previously applied cooling program.

\begin{figure*}
  \centering 
 {\includegraphics[width=3in]{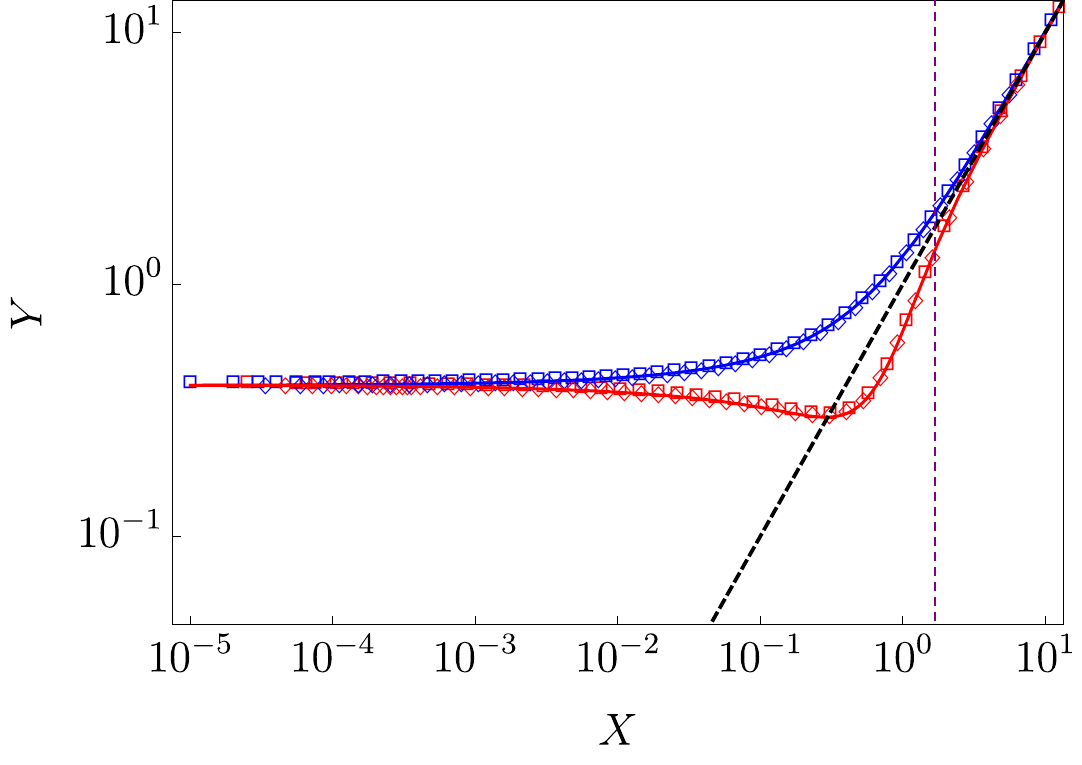} 
 \includegraphics[width=3in]{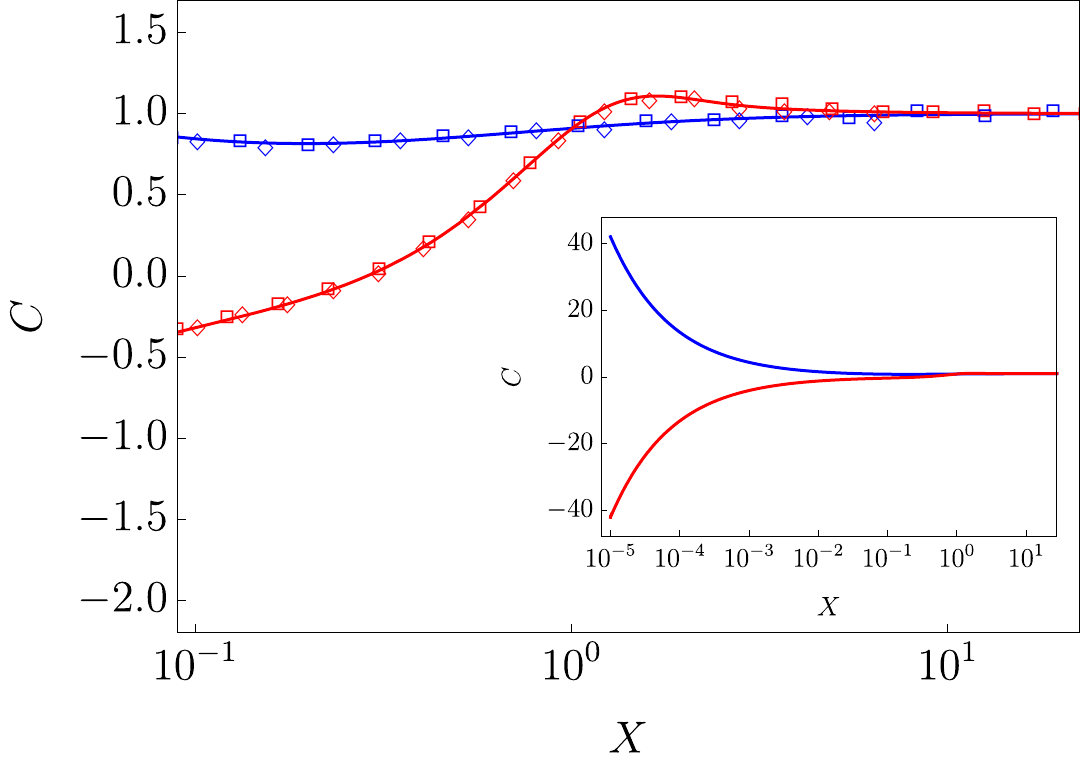}}
  \caption{Hysteresis cycle in the nonlinear molecular fluid.  
Both panels are the transposition to case of the nonlinear molecular fluid of those in Fig.~\ref{hysteresis-gran} for the granular gas, with the same values of the cooling and heating rate $r$ in dimensionless variables. For the molecular fluid, the numerical data {(symbols)} corresponds to the simulation of the Langevin equation \eqref{langevin}, while the theoretical curves {(solid lines)} correspond to the numerical integration of Eqs.~\eqref{ODESinnerc} and \eqref{ODESinnerh}. {On the left panel, the vertical line marks the bath
temperature $X_g$ at which the heat capacity reaches its maximum in the reheating program---see right panel. Therein,} 
%Other parameters employed are $\gamma = 0.1$ and $d=3$. 
an additional inset shows the anomalous behavior of the apparent heat capacity in the cooling process, which is discussed in the text.}
  \label{hysteresis-mol}
\end{figure*}

Figure~\ref{hysteresis-mol} is the transposition of Fig.~\ref{hysteresis-gran} to the case of the nonlinear molecular fluid. Its left panel shows both the numerical simulations of the Langevin equation and the boundary layer solution for a full hysteresis cycle. Similarly to the granular gas case, our boundary layer solution captures very well the simulation data. On the right panel, the behavior of the associated apparent heat capacity of the molecular fluid, ${C=}d\theta/d\theta_{\st}=dY/dX$ is displayed. In the reheating curve, the typical maximum that may be used to define a glass transition temperature is neatly observed. Interestingly, in the cooling curve, an anomalous behavior emerges, the apparent heat capacity increases instead of going to a constant. This anomalous behavior {stems from the singular behavior for small $X$ of the dynamic equation~\eqref{ODESinnerc-Y} for $Y$ in the cooling protocol, and it} is better discerned in the inset---which shows a zoom of the very low temperatures region. {Specifically,} one has that
\begin{align}\label{ODESinnerc-approx}
C=\frac{dY}{dX} &\sim  2\gamma(d+2)\frac{(Y^{\Frz})^2}{X^{1/2}} (1+a_2^{\Frz}), \quad X\ll 1 ,
\end{align}
which diverges as $X^{-1/2}$. This has to be contrasted with the behavior for the granular gas: {Eq.~\eqref{eq:C-low-temp-gran} tells us that $C$ goes to a constant for the granular gas}---consistently with the {results} reported in Fig.~\ref{hysteresis-gran}. 

Finally, our molecular fluid also presents an universal curve when reheated from different frozen states. A regular perturbation theory, once more analogous to that carried out before for the granular gas, gives
 \begin{subequations}
 \label{eq:perturbative-mol-heat-o1}
     \begin{align}
         \theta &=\theta_{\st} -  \frac{r_h}{2\theta_{\st}^{1/2}}\, \frac{1+\gamma (d+6)}{\left[1 + \gamma (d+4)\right]^2 - 2\gamma^2
   (d+4)}, \label{eq:perturbative-mol-heat-o1-a}
         \\
         a_2 &= +\frac{r_h}{\theta_{\st}^{3/2}}\frac{\gamma}{\left[1 + \gamma (d+4)\right]^2 - 2\gamma^2 (d+4)},
     \end{align}
 \end{subequations}
neglecting $O(r_h^2)$ terms. These expressions are obtained from Eqs.~\eqref{eq:perturbative-gran-o1} by exchanging $r_c\leftrightarrow -r_h$. They are valid for {$\theta_{\st}\gtrsim r_h^{2/3}$, i.e., for high enough temperatures such that the system is close to the instantaneous equilibrium curve.}

\begin{figure}
	\centering
	\includegraphics[width=0.98\linewidth]{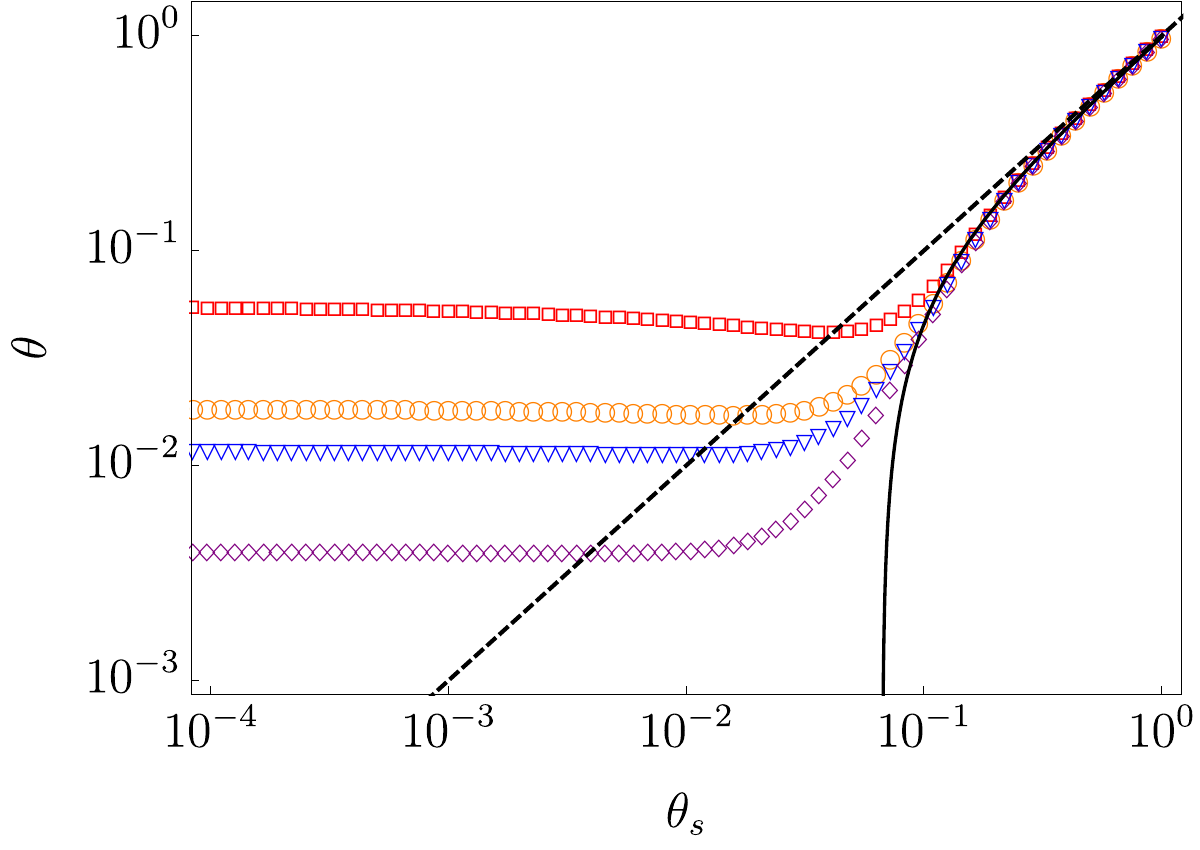} 
 \caption{{Hysteresis cycles for the nonlinear fluid upon reheating with rate $r_h$ from the frozen states corresponding to different cooling rates $r_c$. All curves correspond to $r_h = 0.01$, whereas the cooling rates are:} (color, $r_c$) = (red, $0.05$), (orange, $0.01$), (blue, $0.005$), and (purple, $0.001$). Symbols are simulation results{, and} the  solid curve corresponds to the {perturbative expression~\eqref{eq:perturbative-mol-heat-o1-a} for the normal curve.} 
 %Other parameters are $\gamma = 0.1$ and $d =3$.
 }
	\label{univ-heating-molec}
\end{figure}

\begin{figure}
	\centering
	\includegraphics[width=0.98\linewidth]{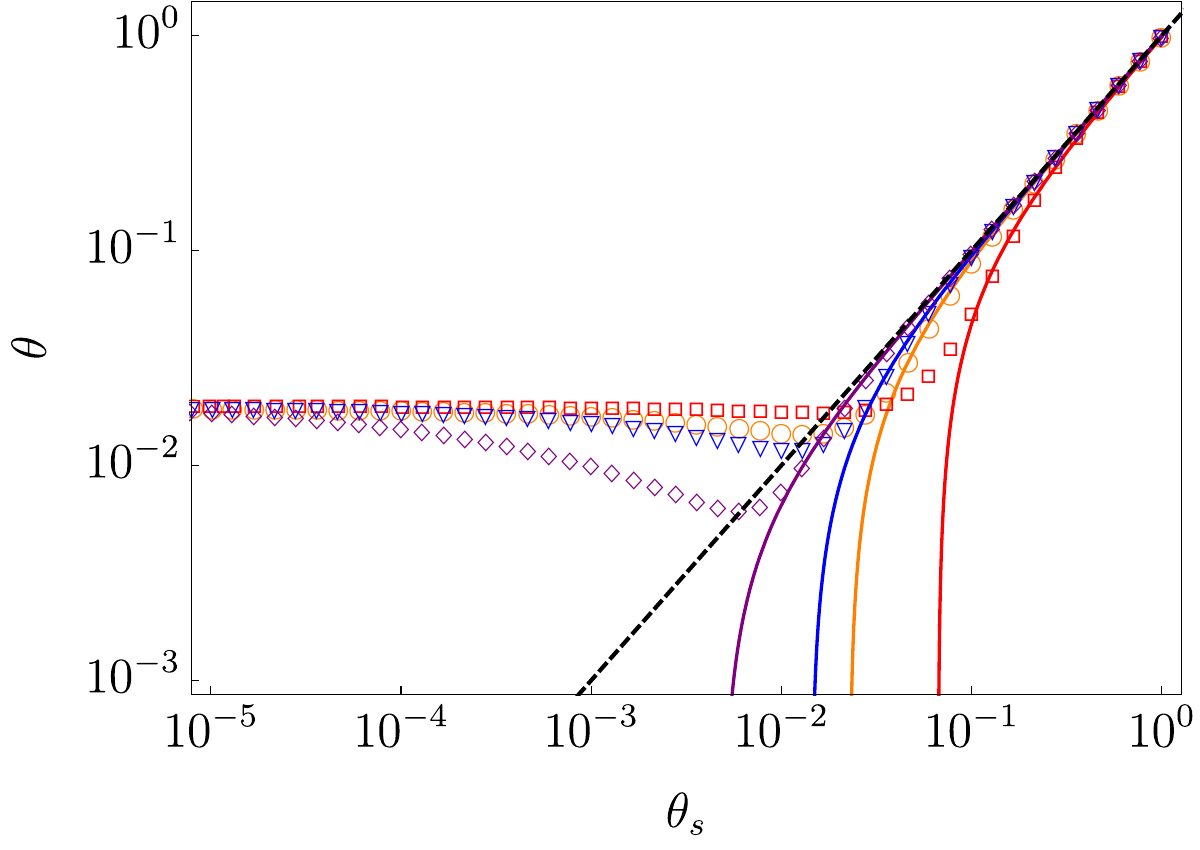} 
 \caption{{Hysteresis cycles for the nonlinear molecular fluid upon reheating with different rates $r_h$ after a common cooling protocol with rate $r_c$. Specifically, the plotted data correspond to $r_c=0.01$ and the reheating rates:  (color, $r_h$) = (red, $0.05$), (orange, $0.1$), (blue, $0.005$) and (purple, $0.001$). As in Fig.~\ref{univ-heating-molec}, symbols correspond to simulation data, whereas the solid curves correspond to Eq.~\eqref{eq:perturbative-mol-heat-o1-a}. 
 %Once more, $\gamma = 0.1$ and $d =3$.
 }}
	\label{further-normal-curve-molecular}
\end{figure}

{When} the system is reheated from  different initial frozen states with kinetic temperatures $\theta^{\Frz} = Y^{\Frz}r_c^{2/3}$,  obtained from previously applied cooling programs with different rates $r_c$, we expect the kinetic temperature to tend towards {the normal curve~\eqref{eq:perturbative-mol-heat-o1-a}} as $\theta_{\st}$ increases. This entails the behavior shown in Figs.~\ref{univ-heating-molec} {and~\ref{further-normal-curve-molecular}}: all the heating curves, independently of the previous cooling rate $r_c$, overshoot the equilibrium curve {to approach the universal normal curve \eqref{eq:perturbative-mol-heat-o1-a}} when being reheated with rate $r_h$. {The discussion on the magnitude of the dimple shown by the reheating curves, which is more marked for higher cooling rates, is completely analogous to that for the granular gas---we thus do not repeat it here.}

\begin{figure}
	\centering	\includegraphics[width=0.95\linewidth]{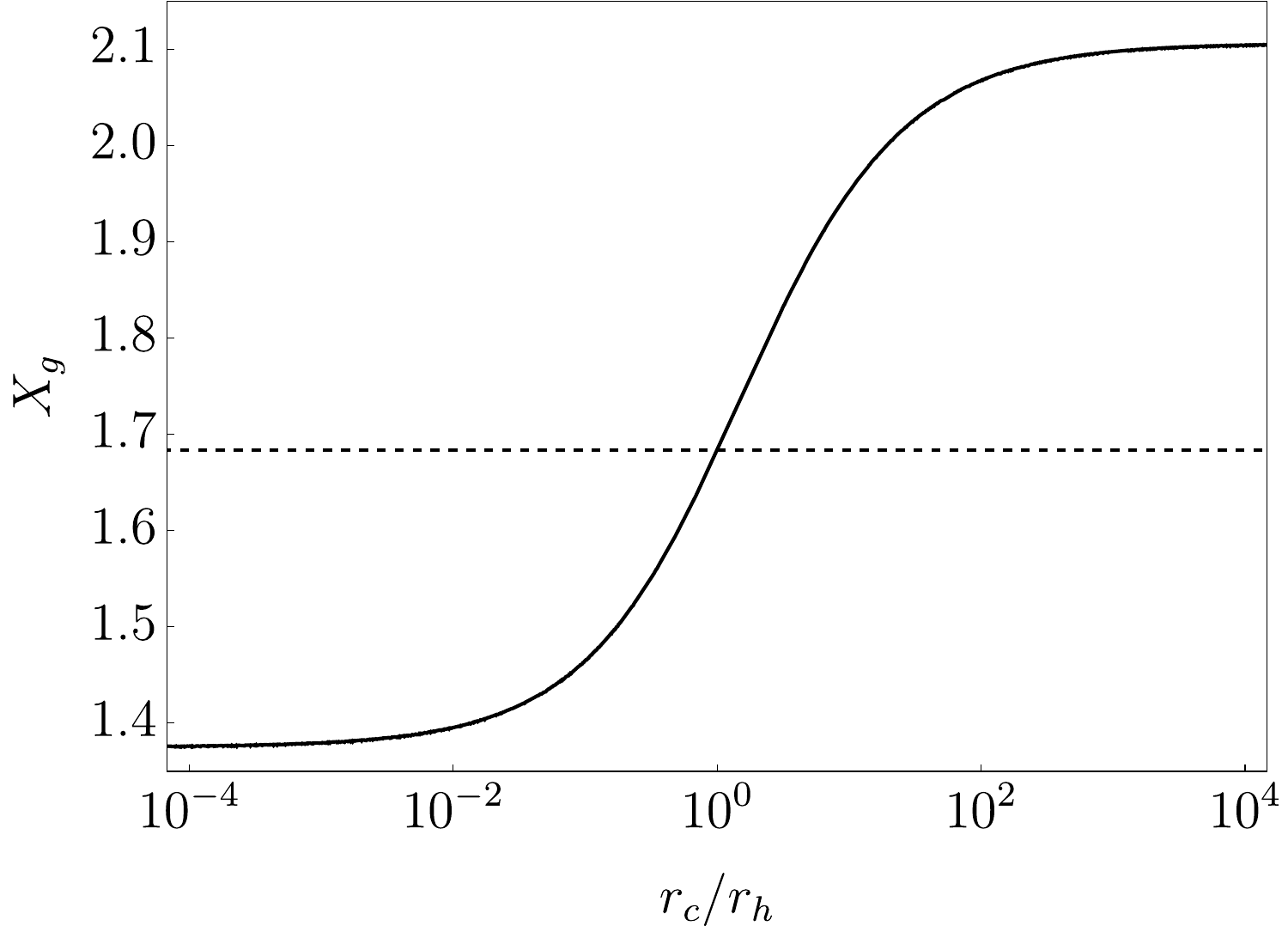} 
 \caption{{Scaled glass transition temperature $X_g=r_h^{-2/3}\theta_g$ for the molecular fluid as a function of the ratio $r_c/r_h$. Similarly to Fig.~\ref{transition-temperature} for the granular gas, we only plot our theoretical prediction for $X_g$, stemming from the numerical integration of the evolution equations in scaled variables, and the dashed horizontal line marks the value for $r_c = r_h$.}}
	\label{transition-temperature-mol}
\end{figure}

{Figure~\ref{transition-temperature-mol} puts forward our theoretical prediction for $X_g$, the scaled glass transition temperature for the molecular fluid, as a function of the ratio of the cooling and heating rates $r_c/r_h$. Again, the glass transition temperature is defined as that at which the apparent heat capacity reaches its maximum. The behavior is once more analogous to the one observed previously for the granular gas, $X_g$ remains of the order of unity over the whole range of $r_c/r_h$. Therefore, the glass transition temperature $\theta_g=r_h^{2/3}X_g$ is also proportional to $r_h^{2/3}$ in the molecular fluid. }

\section{Conclusions} \label{section6}

We have investigated the emergence of a {kinetic} glass transition in two basic fluid models: a granular gas of smooth hard spheres and a molecular fluid with nonlinear drag force. The two systems are very different from a fundamental point of view. One the one hand, collisions in the granular gas are inelastic, and thus its VDF is always non-Gaussian and the system is intrinsically out-of-equilibrium, tending eventually to a NESS if an energy injection mechanism is introduced. On the other hand, collisions are elastic in the molecular fluid and the system approaches equilibrium, with a Maxwellian VDF, in the long time limit. 

In both cases, our analysis have been carried out within the first Sonine approximation of the relevant evolution equation for the VDF: the inelastic Boltzmann equation for the granular gas, the Fokker-Planck equation for the molecular fluid with nonlinear drag. Therein, the evolution equation of the kinetic temperature---basically, the average kinetic energy---is found to be coupled with that of the excess kurtosis. {In turn,} the evolution equation of the excess kurtosis is coupled with higher-order cumulants. {Still, only the excess kurtosis is kept within the first Sonine approximation, higher-order cumulants are neglected because they are assumed to be small.} 

{
In this paper, we have focused on the time evolution of these systems when the bath temperature is decreased to very low values, and afterwards reheated. Despite the profound differences between granular gases and molecular fluids, both systems share some striking similarities in their dynamical behavior. These similarities stem from the algebraic divergence of the characteristic relaxation time, specifically as $T_{\st}^{-1/2}$, for low enough bath temperature. 

The common divergent behavior of the relaxation time in granular gases and nonlinear molecular fluids entails that the frozen values of the kinetic temperature and the excess kurtosis share the same scalings with the cooling rate. Moreover, these scalings can be intuitively understood with simple physical arguments. By defining an effective timescale, which basically measures the number of relaxation times up to the final time of the cooling process, the bath temperature at which the system departs from the instantaneous stationary curve and gets frozen---the so-called fictive temperature---is accurately predicted. 
}

{ Our mathematical approach to the dynamical problem employs a perturbation theory that assumes that the cooling rate is a small parameter. From a mathematical standpoint, the breakdown of the resulting regular perturbative series for low bath temperatures signals the necessity of introducing tools from singular perturbation theory, such as boundary layer techniques. From a physical standpoint, it is interesting to remark that the temperature range at which the regular perturbative approach breaks down coincides with the fictive temperature predicted by intuitive arguments.

Our intuitive physical arguments and our detailed mathematical approach predict, for both granular gases and nonlinear molecular fluids, that the kinetic temperature and the excess kurtosis deviate from their instantaneous stationary curves at low bath temperatures, getting frozen in a value  (i) scaling as $r_c^{2/3}$ for the kinetic temperature and (ii) independent of $r_c$ for the excess kurtosis. In addition, these theoretical predictions have been confirmed by our numerical results: DSMC simulations of the inelastic Boltzmann equation for the granular gas, and numerical integration of the nonlinear Langevin equation for the molecular fluid.
}

A key point of our approach is the evolution equations becoming independent of the cooling rate when they are written in terms of scaled variables, well-suited for our boundary layer treatment of the problem. This {is the mathematical reason for the frozen value of the excess kurtosis being independent of $r_c$. This independence suggests that the complete VDF, beyond the first Sonine approximation employed in the paper, is universal---in the sense of being independent of the cooling rate in scaled variables. We have numerically confirmed this expectation in the numerical simulations of both the granular gas and the nonlinear molecular fluid.}

Moreover, when the system is reheated from this frozen state with the same rate, the independence on the rate, i.e., the above universality, extends to the whole dynamical evolution. This entails that the observed hysteresis when the systems are submitted to a thermal cycle---first cooling, followed by reheating---is also universal, independent of the rate of variation of the bath temperature. Once more, this theoretical prediction {has been} confirmed by numerical simulations of both systems, and an excellent agreement between the numerical and the theoretical curves {has} been found.

Another interesting feature of both systems is their tendency to a unique normal curve upon reheating, independent of the previous cooling program. This behavior has been theoretically predicted for Markovian systems obeying master equations~\cite{brey_normal_1993}, and observed in a variety of simple models for glasses and dense granular systems~\cite{brey_dynamical_1994,brey_dynamical_1994-1,prados_hysteresis_2000,prados_glasslike_2001}. { From a mathematical point of view,} it is this tendency to approach the normal curve that explains the { hysteresis cycles observed upon reheating, including the} overshoot of the instantaneous stationary curve~\footnote{The NESS curve for the granular gas and the equilibrium curve for the molecular fluid.} $\theta=\theta_{\st}$  upon reheating: the normal curve lies below $\theta=\theta_{\st}$ whereas the cooling curves lie above $\theta=\theta_{\st}$.

{ Physically, the hysteresis cycles upon reheating are understood by taking into account that the kinetic temperature in both systems always lags behind the bath temperature---i.e. the former is not able to keep up with the latter for low bath temperatures, due to the divergence of the cooling rate. On the one hand, for the cooling process, this picture leads to the departure of the kinetic temperature from the instantaneous NESS curve $\theta(t)=\theta_{\st}(t)$, with $\theta>\theta_{\st}$. In terms of the apparent heat capacity, this entails the ``free cooling'' behavior of the kinetic temperature, which provides the nonexponential, algebraic, relaxation functions given by Haff's law---for the granular gas---and by the LLNES---for the nonlinear fluid. On the other hand, for the reheating process, this lagging behind the bath temperature also explains the overshooting of the instantaneous NESS curve: for low bath temperatures, the kinetic temperature continues to decrease. Therefore, the kinetic temperature touches the instantaneous NESS curve at a bath temperature for which the relaxation time is still very large, and the kinetic temperature is not able to keep up with the rate of change of the bath temperature---which leads to the overshoot.}

In the granular gas, the values of the excess kurtosis at the frozen state are very close to that of the HCS: this hints at the frozen state being strongly related with the HCS. In the nonlinear molecular fluid, the value of the excess kurtosis are further from that at the LLNES, so the relation between the frozen state and the LLNES is less clear. Still, it seems that both the HCS for the granular gas and the LLNES for the nonlinear fluid play the role of a reference state for the cooling protocol---a first step in this direction is provided in Appendix~\ref{sec:appendix-2}, although this point certainly deserves further investigation. 

The universality of the frozen state, in the sense of its independence of $r_c$ in scaled variables, is an appealing feature of the {kinetic} glass transition found in this work---both for the smooth granular gas and the molecular fluid with (quadratic) nonlinear drag. The possible extension of this property to other systems, for example rough granular fluids~\cite{brilliantov_translations_2007,kremer_transport_2014,torrente_large_2019}, molecular fluids with more complex nonlinearities~\cite{klimontovich_nonlinear_1994,lindner_diffusion_2007,casado-pascual_directed_2018,goychuk_nonequilibrium_2021,patron_strong_2021}, or binary mixtures~\cite{serero_hydrodynamics_2006,khalil_homogeneous_2014,gomez_gonzalez_mpemba-like_2021,gomez_gonzalez_time-dependent_2021}, is an interesting prospect for future work.

\begin{acknowledgments}
We acknowledge financial support from Grant PID2021-122588NB-I00
funded by MCIN/AEI/10.13039/501100011033/ and by ``ERDF A way of
making Europe''. We also acknowledge financial support from Grant
ProyExcel\_00796 funded by Junta de Andalucía's PAIDI 2020 programme. A.~Patr\'on acknowledges support from the FPU programme through Grant FPU2019-4110.
\end{acknowledgments}

\section*{Data availability}
The Fortran codes employed for generating the data that support the findings of this study, together with the Mathematica notebooks employed for producing the figures presented in the paper, are openly available in the~\href{https://github.com/fine-group-us/laboratory-glass-transition}{GitHub page} of University of Sevilla's FINE research group.

\appendix

\section{\label{sec:appendix-2} Glass transition for different cooling programs}

\begin{figure*}
  \centering 
 {\includegraphics[width=3in]{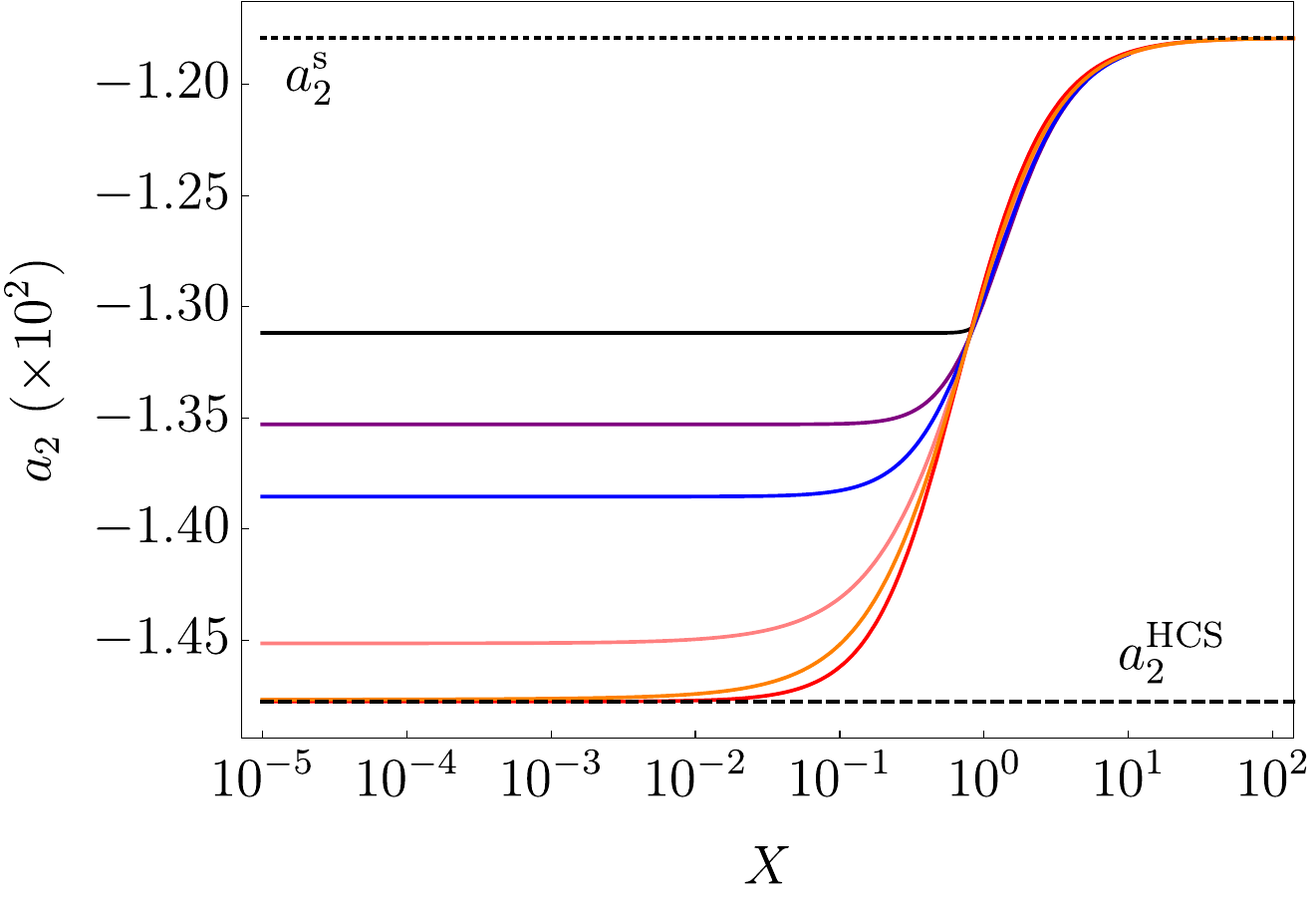} 
 \includegraphics[width=3.59in]{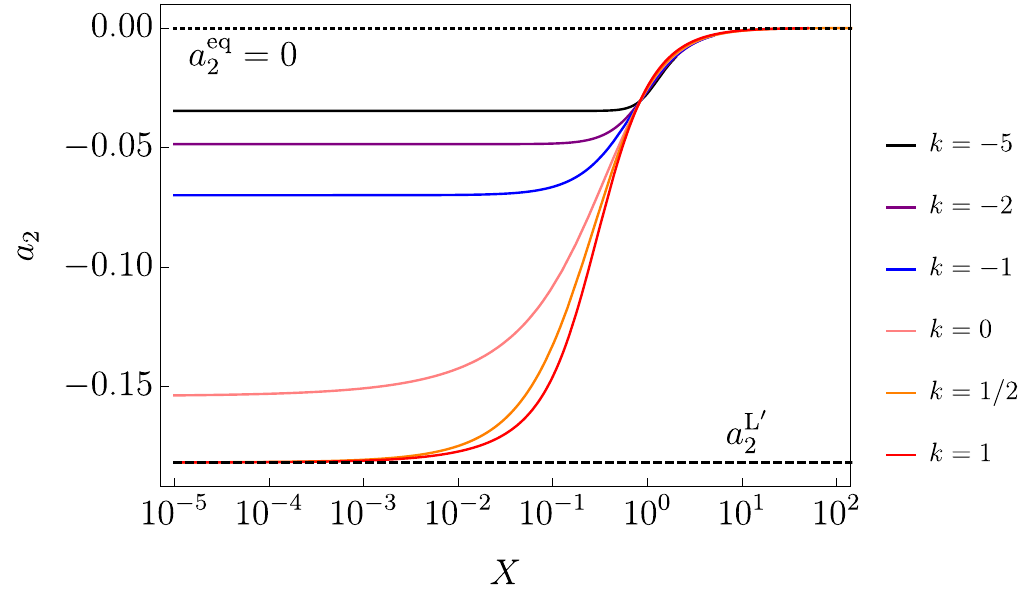}}
  \caption{Excess kurtosis $a_2$ as a function of the scaled bath temperature $X$ for {the family of cooling programs}~\eqref{eq:general-cooling} with different values of $k$. {The solid lines correspond to the numerical integration of the evolution equations in the first Sonine approximation.} (Left) behavior found in a granular {gas}. (Right) Behavior found in a molecular fluid. In both cases, a dimensionless cooling rate $r_c=0.1$ has been employed. On the left (right) panel, {(i)} the  dotted line represents the value of the excess kurtosis at the NESS $a_2^{\st}$ (equilibrium value of the excess kurtosis $a_2^{\text{eq}}=0$), from which all the curves depart for large $X$, {and (ii)} the  dashed line accounts for the value at the HCS (LLNES) within the first Sonine approximation.}
  \label{varying-k-a2}
\end{figure*}
Throughout this work, we have employed linear cooling programs in order to study the emergence of glassy behavior in both molecular fluids and granular gases. Let us now consider the following {general} family of cooling protocols,
\begin{equation}\label{eq:general-cooling}
    \frac{d\theta_{\st}}{dt} = -r_c \theta_{\st}^{k}, 
\end{equation}
with $k$ being a real number. Notice that the $k=0$ case reduces to the already studied linear cooling {protocol}. 

We still consider that the cooling is slow, in the sense that $r_c\ll 1$. {Following the same approach as in Sec.~\ref{sec:why-kinetic-glass-transition} for the linear cooling program, we introduce the effective timescale
\begin{equation}
    s = \int_{t}^{t_0}dt' \tau^{-1}(\theta_{\st}(t')) = \frac{1}{r_c}\int_0^{\theta_{\st}}d\theta_{\st} \ (\theta_{\st}')^{-k}\tau^{-1}(\theta_{\st}'),
\end{equation}
where we have inserted Eq.~\eqref{eq:general-cooling} into the second integral. Assuming again that $s$ is of the order of unity when the system freezes, we arrive at the following estimation of the fictive temperature $\theta_f$;
\begin{equation}\label{app:theta-f-scaling-with-k}
    \theta_f = \left[\left(\frac{3}{2}-k\right)\frac{r_c}{c}\right]^{\frac{2}{3-2k}}, \quad \theta_{\text{Frz}} \propto r_c^{\frac{2}{3-2k}},
\end{equation}
where we have taken into account that the characteristic relaxation timescale $\tau$ is still given by Eq.~\eqref{eq:tau-def}, regardless of the cooling program employed. Thus, the above entails that the system will present a kinetic glass transition as long as $k<k_{\text{crit}}=3/2$. Eq.~\eqref{app:theta-f-scaling-with-k} generalizes the power law behavior $r_c^{2/3}$ found in the main text for $k=0$.
}

{Now, in order to corroborate the scaling found in Eq.~\eqref{app:theta-f-scaling-with-k}, we follow }{a regular perturbation approach similar to} the ones employed {in the main text} for both the granular gas and the nonlinear fluid. {The} solution to the lowest order corresponds again to the {instantaneous} stationary solutions $\theta^{(0)} = \theta_{\st}$, $a_2^{(0)}=a_2^{\st}$. The first-order $O(r_c)$ corrections  {are} provided by the equations
\begin{subequations}
    \begin{align}
        -\theta_{\st}^k &= \theta_{\st}^{1/2}\left\{c_1 \theta^{(1)} + c_2\theta_{\st}a_2^{(1)}\right\},
        \\
        0 &=c_3 \frac{\theta^{(1)}}{\theta_{\st}^{1/2}} + c_4\theta_{\st}^{1/2}a_2^{(1)},
    \end{align}
\end{subequations}
with $c_i, \ i=1,..,4$ being constants that depend on the parameters of the specific system of concern. These equations entail the scalings
\begin{equation}
\label{eq_scaled-varying-k}
    \theta^{(1)} \propto \theta_{\st}^{k-\frac{1}{2}}, \quad a_2^{(1)} \propto \theta_{\st}^{k-\frac{3}{2}}.
\end{equation}
which imply that $\theta^{(1)}\ll\theta^{(0)}=\theta_{\st}$, $a_2^{(1)}\ll a_2^{(0)}=a_2^{\st}$ when $k > k_{\crit}= 3/2$ . Therefore, for $k>k_{\crit}$ {a kinetic glass transition is presented neither by the granular gas nor by the nonlinear fluid. The}  cooling is so slow for $k>k_{\crit}$ that both systems remain basically over the {instantaneous} stationary curve $\{\theta=\theta_{\st}, a_2=a_2^{\st}\}$ for all bath temperatures~\footnote{The existence of a critical value of $k$ for the emergence of a glass transition has already been reported in simple models~\cite{brey_residual_1991,brey_dynamical_1994}.}

Let us now consider the case $k\leq k_{\crit}$. In this case, the regular perturbation approach breaks down for low enough bath temperatures, which marks the onset of the {kinetic} glass transition. Our regular perturbation approach ceases to be valid when the $O(1)$ terms become comparable with the $O(r_c)$ ones, thus implying
\begin{equation}
    \theta_{\st} = O\left(r_c^{\frac{2}{3-2k}}\right),
    \label{eq:thetast-func-k}
\end{equation}
{which, consistently with our discussion above, only makes sense for $k<k_{\crit}$, it diverges for $k>k_{\crit}$.} Equation~\eqref{eq:thetast-func-k} entails that we expect that the kinetic temperature at the frozen state scales as $\theta^{\Frz}\propto r_c^{\frac{2}{3-2k}}$, { consistently with Eq.~\eqref{app:theta-f-scaling-with-k}.} Interestingly, regardless of the choice of $k$, the frozen state is still universal, in the sense that it is independent of the cooling rate $r_c$, since
\begin{equation}
    a_2 - a_2^{\st}\sim a_2^{(1)} \propto r_c \,\theta_{\st}^{k-\frac{3}{2}} = O(1).
\end{equation}

As in the main text, the above scaling relations suggest the introduction of scaled variables
\begin{equation}
    Y \equiv r_c^{-\frac{2}{3-2k}}\theta , \quad X \equiv r_c^{-\frac{2}{3-2k}}\theta_{\st} .
\end{equation}
In terms of them, the dynamic equations for the cooling protocol become $r_c$-independent. The same applies for a reheating program with $r_h=r_c$ from the frozen state. We remark that the evolution equations for the scaled variables $(Y,a_2)$ in both systems are the same as the ones we have written in the main text, with the only change $ d/dX \leftrightarrow  X^k d/dX$ on their lhs.

Figure~\ref{varying-k-a2} shows the evolution of the excess kurtosis $a_2$ towards its frozen state in a cooling program with rate $r_c$ for different values of $k$ in Eq.~\eqref{eq:general-cooling}, for both the granular gas and the nonlinear molecular fluid. {The} excess kurtosis follows a similar trend: on the one hand, for $k\rightarrow -\infty$, the time window over which $\theta_{\st}$ decays towards zero becomes infinitely small, and thus the excess kurtosis does not have time to deviate from its stationary state value and is approximately constant for all $X$.  On the other hand, as the value of $k$ is increased, the time window to relax also increases. The limiting case $k = k_{\crit}$ constitutes the ultimate balance between a sufficiently wide time window to relax, and a fast enough relaxation protocol such that $\theta$ deviates from the $\theta = \theta_{\st}$ behavior. 

It is worth noting that, for $1/2 \leq k < k_{\text{crit}}$, the excess kurtosis tends to the value over the HCS---for the granular gas--- and the LLNES---for the nonlinear molecular fluid. The lower bound $k = 1/2$ corresponds to the value above which the deviations from the $\theta = \theta_{\st}$ line become significantly small, but still allowing for the kurtosis to evolve towards the frozen state, as Eq.~\eqref{eq_scaled-varying-k} states. Since we are showing the numerical integration of the evolution equations in the first Sonine approximation, these limit values of the excess kurtosis correspond to their theoretical estimates in this framework. For the granular gas, this is given by Eq.~\eqref{eq:a2hcs}, which is quite accurate due to its smallness. For the nonlinear fluid, the first Sonine approximation gives ${a_2^{\LLNES}}^{\prime}=-2/(d+8)$~\cite{patron_strong_2021}, which is quite different from its exact value in Eq.~\eqref{eq:a2-L}---this is reasonable, since the deviations from the Gaussian are much larger in the LLNES than in the HCS.

The above discussion hints at the frozen state corresponding  to the HCS and the LLNES for the granular gas and the nonlinear molecular fluid, respectively, {for $1/2 \leq k < k_{\text{crit}}$. This means that the two model systems,} either the granular gas or the nonlinear molecular fluid, reach the corresponding nonequilibrium state, either the HCS or the LLNES, over a time window of the order of $r_c^{-1}$ when {cooled down} with a program for which $1/2 \leq k < k_{\text{crit}}$. The latter suggests useful applications in optimal control~\cite{aurell_optimal_2011,prados_optimizing_2021,guery-odelin_driving_2023,blaber_optimal_2023} and also within the study of nonequilibrium effects, as previous work on both systems shows that both the HCS and the LLNES are responsible for the emergence of a plethora of nonequilibrium phenomena, such as the Mpemba and Kovacs effects~\cite{prados_kovacs-like_2014,trizac_memory_2014,lasanta_when_2017,patron_strong_2021,patron_nonequilibrium_2023,patron_nonequilibrium_2023}.

\section{\label{app:linear-relaxation-time} Relaxation time of the granular gas}

{
In Ref.~\cite{sanchez-rey_linear_2021}, the linear relaxation to the NESS of the uniformly heated granular gas was investigated in detail. The relaxation of the system from the NESS corresponding to a value of the driving intensity $\xi+\delta\xi$ to the NESS corresponding to a driving intensity $\xi$ was considered. 

Therefore, the granular (or kinetic) temperature evolves from the initial value $\theta_{\st}+\delta\theta_{\st}$ to the final value $\theta_{\st}$. The relaxation function for the granular temperature can be defined as,
\begin{equation}
    \phi_{\theta}(t)\equiv \frac{\theta(t)-\theta_{\st}}{\delta\theta_{\st}},
\end{equation}
which is normalised in the usual way, $\phi_{\theta}(t=0)=1$. In particular, this relaxation function was shown to have the following form:
\begin{equation}
    \phi_{\theta}(t)=a_+ e^{-\lambda_+ t}+a_- e^{-\lambda_{-} t},
\end{equation}
where, due to the normalization, $a_-=1-a_+$. Both $\lambda_+$ and $\lambda_+$ are proportional to $\theta_{\st}^{1/2}$
\begin{equation}
\lambda_{\pm}=c_{\pm}\theta_{\st}^{1/2},
\end{equation}
with
\begin{subequations}
  \begin{align}
    c_+&=\frac{3}{2}+\frac{9}{32}\frac{1+4B}{4B-3}a_2^{\st}+O(a_2^{\st})^2, \\
    c_-&=2B-\frac{9}{4(4B-3)}a_2^{\st}+O(a_2^{\st})^2,
\end{align}  
\end{subequations}
and 
\begin{equation}
    a_+=1+\frac{9}{4(4B-3)}a_2^{\st}+O(a_2^{\st})^2.
\end{equation}

The characteristic relaxation time of the relaxation is given by
\begin{equation}
    \tau \equiv \int_0^{\infty} dt\, \phi_{\theta}(t)=\frac{a_+}{\lambda_+}+\frac{a_-}{\lambda_{-}} .
\end{equation}
Since both $\lambda_+$ and $\lambda_-$ are proportional to $\theta_{\st}^{1/2}$, and both $a_+$ and $a_-$ are independent of the bath temperature, we have the scaling in Eq.~\eqref{eq:tau-def} of the main text,
\begin{equation}
    \tau=\frac{1}{c} \theta_{\st}^{-1/2}, \quad \frac{1}{c}=\frac{a_+}{c_+}+\frac{a_-}{c_-}.
\end{equation}
In practice, it is straightforward to check numerically that $a_{+} \approx 1 \gg a_{-}$ regardless of the value of $\alpha$, such that $c\approx c_{+} \approx 3/2$, which provides the value $X_f = 1$ for the fictive temperature that it is depicted Fig.~\ref{univ-cooling-gran}. Thus, we may roughly state that the fictive temperature is independent of the restitution coefficient.}

%

%\vspace{1ex}
%%

\section{\label{app:reg-granular-gas} Regular perturbation theory for the granular gas}

{
In order to find an approximate solution of the evolution equations for cooling~\eqref{granularSonine-v2}, we introduce the regular perturbation series in powers of the cooling rate,
\begin{align}
\label{app:pert-gran}
    \theta &= \theta^{(0)} + r_c \; \theta^{(1)} + O(r_c^2), \nonumber \\
    a_2& = a_2^{(0)} + r_c \; a_2^{(1)} + O(r_c^2).
\end{align}
Eqs.~\eqref{app:pert-gran} are inserted into Eq.~\eqref{granularSonine-v2}, in which we subsequently equal the terms with the same power of $r_c$ and solve for $\{\theta^{(k)},a_2^{(k)}\}$, $k=0,1,\ldots$. At the lowest order, $O(r_c^0)$, i.e., for terms independent of $r_c$, we get
\begin{equation}
		\label{app:r0-sol}
		\theta^{(0)} = \theta_{\st}, \quad a_2^{(0)} = a_2^{\st},
\end{equation}
which corresponds to the instantaneous NESS curve. At the first order, $O(r_c)$, i.e., for terms linear in $r_c$, we get
\begin{subequations} 
\label{app:order1}
	\begin{align}
			\theta^{(1)} &= \frac{2}{3} \theta_{\st}^{-1/2}  \left[1+\frac{3}{16} \; a_2^{\st} \left( 1+\frac{1}{B} \right) \right]^{-1},
			\\
			a_2 ^{(1)} &= -\frac{a_2^{\st}}{B} \theta_{\st}^{-3/2} \left[1+\frac{3}{16} \; a_2^{\st} \left( 1+\frac{1}{B} \right) \right]^{-1}.
	\end{align}
\end{subequations}
Putting together Eqs.~\eqref{app:r0-sol} and \eqref{app:order1}, we obtain the regular perturbative expression~\eqref{outer-gran-cooling} in the main text. 
}

\section{\label{app:boundary-layer-granular} Uniform solution in the boundary layer approach}

{
In this appendix, we aim to derive an approximate expression for the temperature and the excess kurtosis for the cooling process, valid over the whole bath temperature range. For high enough bath temperatures, $\theta_{\st}\gg r_c^{2/3}$, we have the ``outer" expansion in Eq.~\eqref{outer-gran-cooling}. For low enough bath temperatures, $\theta_{\st}=O(r_c^{2/3})$, we have the boundary layer system~\eqref{eq:evolscaled} for the scaled variables. In boundary layer perturbation theory, the lowest-order perturbative solution---knowns as the ``uniform" solution---is constructed as the sum of the lowest-order outer and inner solutions, minus the common behavior found in an intermediate matching region~\cite{bender_advanced_1999}. Below we derive such a uniform solution for the cooling protocol.

We denote the outer solution at the lowest order by $(\theta_O,a_{2,O})$. Equation~\eqref{outer-gran-cooling} tells us that
\begin{equation}
\label{app:outer-series}
    \theta_O(\theta_{\st})=\theta_{\st}, \quad a_{2,O}(\theta_{\st})=a_{2}^{\st},
\end{equation}
which is the instantaneous NESS curve. Now, let us seek the solution of the inner problem at the lowest order, which we denote by $(Y_I(X),a_{2,I}(X))$. To obtain it, we solve Eq.~\eqref{eq:evolscaled} with the boundary conditions
\begin{equation}
    \lim_{X\to\infty}Y_I(X) = \infty, \quad \lim_{X\to\infty} a_{2,I}(X) = a_2^{\st},
    \label{app:boundary-infty}
\end{equation}
which correspond to the limit as $r_c\to 0$ in Eq.~\eqref{eq:boundary}. Therefore, $(Y_I(X),a_{2,I}(X))$ does not depend on $r_c$, since neither the evolution equations~\eqref{eq:evolscaled} nor the boundary conditions~\eqref{app:boundary-infty} depend on $r_c$. 

Although it is not possible to write $(Y_I(X),a_{2,I}(X))$ in a simple closed form, an asymptotic analysis for large values of $X$ gives
\begin{equation}
    Y_I(X)\sim X, \qquad a_{2,I}(X)\sim a_2^{\st}, \quad X\gg 1,
    \label{app:inner-asymp}
\end{equation}
which is consistent with the tendency of the DSMC data in Fig.~\ref{univ-cooling-gran} to the instantaneous NESS curve for large $X$. By comparing Eqs.~\eqref{app:outer-series} and \eqref{app:inner-asymp}, we obtain the common behavior
\begin{equation}
    \theta_c (\theta_{\st})=\theta_{\st}, \qquad a_{2,c}(\theta_{\st})=a_2^{\st}, \quad \theta_{\st}\ll 1, \; X\gg 1,
    \label{app:common-behavior}
\end{equation}
or, $r_c^{2/3}\ll \theta_{\st}\ll 1$. This is the region at which the outer and inner solution match, the uniform solution is built as~\cite{bender_advanced_1999}
\begin{align}
    \theta^{\BL}(\theta_{\st})&=\theta_O(\theta_{\st})+r_c^{2/3}Y_I(X=r_c^{-2/3}\theta_{\st})-\theta_c(\theta_{\st}),\\
    a_{2}^{\BL}(\theta_{\st})&=a_{2,O}(\theta_{\st})+a_{2,I}(X=r_c^{-2/3}\theta_{\st})-a_{2,c}(\theta_{\st}),
\end{align}
Since the common behavior~\eqref{app:common-behavior} equals the outer solution~\eqref{app:outer-series}, the range of validity of the inner solution  extends to the whole temperature interval. In other words, the uniform solution coincides with the inner solution:
\begin{align}
    \theta^{\BL}(\theta_{\st})&=r_c^{2/3}Y_I(X=r_c^{-2/3}\theta_{\st}), \\
    a_{2}^{\BL}(\theta_{\st})&=a_{2,I}(X=r_c^{-2/3}\theta_{\st}).
\end{align}
The first equation tells us that $Y^{\BL}\equiv r_c^{-2/3}\theta^{\BL}=Y_I$, i.e., we have the universal behavior in Fig.~\ref{univ-cooling-gran} over the uniform solution---to the lowest order.
}

\section{\label{app:reg-pert-mol-fluid} Regular perturbation theory for the molecular fluid}

Following an  approach similar to that in Sec.~\ref{subsec:pert-mol}, let us decrease the bath temperature by applying the linear cooling program
\begin{align}
\label{cooling-program2}
 \frac{d\theta_{\st}}{dt} = -r_c \; \Rightarrow  \frac{d\theta}{dt}= -r_c \frac{d\theta}{d\theta_{\st}}, \quad 
 \frac{da_2}{dt} = -r_c \frac{da_2}{d\theta_{\st}},
 \end{align}
where $r_c\ll 1$ is the cooling rate.  We employ again the  boundary layer theory~\cite{bender_advanced_1999} to approach the problem. For the outer layer, for which it is expected that $\theta$ does not deviate too much from $\theta_{\st}$, we insert the regular perturbation series
\begin{subequations}\label{regular-perturb-mol}
\begin{align}
    \theta &= \theta^{(0)} + r_c  \theta^{(1)} + O(r_c^2), 
    \\
    a_2 &= a_2^{(0)} + r_c a_2^{(1)} + O(r_c^2),
\end{align}
\end{subequations}
into the evolution equations~\eqref{suppl-eq:evol-eqs-first} and equate terms with the same power of $r_c$. At the lowest order, $O(1)$, one obtains
\begin{subequations}\label{suppl-eq:evol-eqs-first-0}
\begin{align}
    0  = & 2(\theta_{\st}-\theta^{(0)}) + 2\gamma (d+2)\, \theta^{(0)} 
%\nonumber \\ & 
- 2\, \gamma \, (d+2)(1+a_2^{(0)})\frac{[\theta^{(0)}]^{2}}{\theta_{\st}} ,\\
    0  = & 8\gamma \left(1-\frac{\theta^{(0)}}{\theta_{\st}}\right) 
%   \nonumber \\ & 
    -\left[\frac{4\theta_{\st}}{\theta^{(0)}}-8\gamma+4\gamma(d+8)\frac{\theta^{(0)}}{\theta_{\st}} \right]a_2^{(0)},
\end{align}
\end{subequations}
whose solution corresponds to equilibrium:
\begin{equation}
\label{order-0-mol}
    \theta^{(0)} = \theta_{\st}, \quad  a_2^{(0)} = 0.
\end{equation}
The linear terms in $r_c$ obey
\begin{subequations}
    \begin{align}
        -1 &= \theta_{\st}^{1/2}\left\{-2\left[1+\gamma (d+2)\right] \theta^{(1)} -2\gamma (d+2) \theta_{\st}a_2^{(1)}\right\},
        \\
        0 &=2\gamma \frac{\theta^{(1)}}{\theta_{\st}^{1/2}} + \theta_{\st}^{1/2}\left[1 + \gamma (d+6)\right]a_2^{(1)}.
    \end{align}
\end{subequations}
The solution of this system is given by 
\begin{align}
\label{order-1-mol}
    \theta^{(1)} &= \frac{1}{2\theta_{\st}^{1/2}}\frac{1+\gamma (d+6)}{\left[1 + \gamma (d+4)\right]^2 - 2\gamma^2 (d+4)}, \\ a_2^{(1)} &= -\frac{1}{\theta_{\st}^{3/2}}\frac{\gamma}{\left[1 + \gamma (d+4)\right]^2 - 2\gamma^2 (d+4)}.
\end{align}
The regular perturbation expansion~\eqref{eq:perturbative-gran-o1} in the main text is directly obtained by combining Eqs.~\eqref{regular-perturb-mol}, \eqref{order-0-mol} and \eqref{order-1-mol}.

\bibliography{Mi-biblioteca-28-sep-2023}

\end{document}